\documentclass[openacc]{rsproca_new}

\usepackage[capitalise]{cleveref}
\usepackage{subcaption}

\usepackage{mathtools}

\newtheorem{condition}{\bf Condition}[section]

\usepackage{braket}
\graphicspath{{./graphics/}}

\begin{document}

\title{The Small Stellated Dodecahedron Code and Friends}

\author{
J.~Conrad$^1$, C.~Chamberland$^{2}$, N.~P.~Breuckmann$^{3}$, B.~M.~Terhal$^{4,5}$}

\address{$^{1}$ JARA Institute for Quantum Information, RWTH Aachen University, Aachen 52056, Germany\\
$^{2}$ Institute for Quantum Computing and Department of Physics and Astronomy, University of Waterloo, Waterloo, Ontario, N2L 3G1, Canada\\
$^{3}$ Department of Physics and Astronomy, University College London, London WC1E 6BT, UK\\
$^{4}$ QuTech, Delft University of Technology, P.O. Box 5046, 2600 GA Delft, The Netherlands \\
$^{5}$ Institute for Theoretical Nanoelectronics, Forschungszentrum Juelich, D-52425 Juelich, Germany}

\subject{quantum information}


\corres{Barbara M. Terhal\\
\email{bterhal@gmail.com}}
\begin{abstract}
We explore a distance-3 homological CSS quantum code, namely the small stellated dodecahedron code, for dense storage of quantum information and we compare its performance with the distance-3 surface code. The data and ancilla qubits of the small stellated dodecahedron code can be located on the edges resp. vertices of a small stellated dodecahedron, making this code suitable for 3D connectivity. This code encodes 8 logical qubits into 30 physical qubits (plus 22 ancilla qubits for parity check measurements) as compared to 1 logical qubit into 9 physical qubits (plus 8 ancilla qubits) for the surface code. We develop fault-tolerant parity check circuits and a decoder for this code, allowing us to numerically assess the circuit-based pseudo-threshold. \end{abstract}




\begin{fmtext}
\end{fmtext}
\maketitle

\tableofcontents
\section{Introduction}

The popular toric or surface codes are members of a family of topological codes called homological CSS codes \cite{FM:surfacecode, BK:surfacecode, freedman2002z2} which can be obtained from tessellations of $D$-dimensional manifolds. Curvature and topology of these manifolds determine features of these codes. Although a code does not specify a specific physical lay-out or physical distance between qubits, its prescription of which parity checks need to be measured, dictates what high-precision interactions need to be engineered between the physical qubits and ancilla qubits for measuring parity checks. As such, a code based on a tessellation of a 2D flat manifold suits a planar 2D connectivity between qubits, while a three-dimensional representation of a code in terms of a polyhedron could be used as a template of how physical qubits could be placed and connected up in 3D. 

In this paper we continue the exploration of so-called hyperbolic surface codes \cite{BT:hyper, Breuckmann+:hyper} to determine whether such codes, being block codes with high rate, have advantages over the surface code. The work in \cite{BT:hyper} constructed various classes of hyperbolic surface codes based on regular tessellations and numerically examined noise thresholds of these codes when subjected to depolarizing noise (assuming noiseless parity checks). The work in \cite{Breuckmann+:hyper} went one step further by including effective noise in the parity check measurements themselves, focusing uniquely on $\{4,5\}$-hyperbolic surface codes. Ref. \cite{Breuckmann+:hyper} also showed how to do read/write operations using Dehn twists if these block codes are used as a quantum memory. In this paper we focus on one of the smallest and simplest members of the hyperbolic surface code family, namely a code which has a representation as a small stellated dodecahedron. Going beyond the previous work, we examine the performance of the code when all elementary gates and operations, including those in the parity check circuits, are noisy (more details of the circuit level noise model are given in \cref{sec:FaultTolerantSchemes}). 

The interest in the small stellated dodecahedron code is that it can pack logical qubits very densely while, like the $[[9,1,3]]$ surface code, still allowing for plain fault-tolerant parity check measurements in combination with a look-up table decoder. Even denser packings of logical qubits in block stabilizer codes are certainly feasible: there are non-CSS codes such as $[[8,3,3]]$, $[[10,4,3]]$, $[[11,5,3]]$, $[[13,7,3]]$, and $[[14,8,3]]$ codes listed in \cite{Grassl:codetables}. However, one may expect that the construction of fault-tolerant parity check circuits for such codes requires resource-intense methods such as Steane, Shor or Knill error correction, or flag-fault-tolerance methods \cite{CR:flag, CB:flag}. Ref.~\cite{CR:flag} also proposed fault-tolerant circuits for a non-topological $[[15,7,3]]$ Hamming code, using only 17 physical qubits in total: a disadvantage of this code is that the weight of the parity checks is high, namely 8, and in the tally of 17 qubits all parity checks are done using the same ancilla qubit.

We find that for a depolarizing circuit-level noise model the stellated dodecahedron code pays for its dense storage with a pseudo-threshold which is a factor 19 lower than that of the Surface-17 code. Despite this somewhat negative message, the methods developed in this paper lay the groundwork for further exploration of these families of codes.  

We start the paper by recalling the notion of homological CSS codes, illustrating this code construction by a variety of examples in 2D, representable as star polyhedra, as well as a few 3D and 4D codes. In \cref{sec:features} we zoom in on the small stellated dodecahedron code, while we zoom out again in \cref{sec:ParityCheckSchedules} by formalizing the problem of optimally scheduling the entangling gates of parity check circuits of LPDC codes or more specifically hyperbolic surface codes. We apply these techniques in \cref{sec:FaultTolerantSchemes} to the dodecahedron code obtaining fault-tolerant circuits and describing the decoding method. In \cref{Sec:Numerics} we report on the results of our numerical implementation, which includes a direct comparison with the Surface-17 code. We end the paper with a Discussion. 

\section{Homological CSS Codes}
\label{subsec:HomologicalCSS}

Here we briefly review the definition of homological CSS codes. 
We start with a regular tessellation of a $D$-dimensional closed manifold. This defines a \textit{cell complex} composed of \textit{i-cells}, with $i=0,1,..,D$ referring to the cell dimension. 
The $i$-cells span a vector space $C_i = \mathbb{Z}_2^{\dim(C_i)}$ whose elements will be called \textit{i-chains}. Given such a cell complex, one can define a CSS code by associating the $i$-cells with physical qubits, the $(i+1)$-cells with $Z$-checks (i.e. generators of elements in the stabilizer group which only involve Pauli $Z$ operators) and the $(i-1)$-cells with $X$-checks (i.e. generators of elements in the stabilizer group which only involve Pauli $X$ operators). The number of physical qubits of the code is $n=\dim(C_i)$. 
A $Z$-parity check is associated with each $(i+1)$-cell and it takes the parity of the qubits/$i$-cells which form the boundary of the $(i+1)$-cell. Formally, the boundary operator $\partial_{i+1}$ is defined as $\partial_{i+1} : \; C_{i+1} \rightarrow C_{i}$. Similarly, a $X$-parity check is associated with each $(i-1)$-cell so that it takes the $X$-parity of all qubits/$i$-cells which are the co-boundary of the $(i-1)$-cell (that is, which have the $(i-1)$-cell as their boundary).  Formally, the coboundary operator $\delta_{i-1}$ is defined as $\delta_{i-1}=\partial_{i}^T : \; C_{i-1} \rightarrow C_{i}$. The $X$- and $Z$-parity checks commute since the boundary of any $(i+1)$-cell and the co-boundary of any $(i-1)$-cell overlap on an even number of $i$-cells/qubits. 

By the parity check weight of a $X$- or $Z$-parity check we mean the number of qubits on which this parity check acts non-trivially. The logical $Z$ operators (denoted as $\overline{Z}$) of the code are closed $i$-chains which are not the boundary of any collections of $(i+1)$-cells. Similarly, the logical $X$ operators (denoted as $\overline{X}$) are closed $i$-cochains which are not the co-boundary of any collection of $(i-1)$-cells. The number of logical qubits of the code is given by $k=\dim(H_i(\mathbb{Z}_2)) $ where $H_i(\mathbb{Z}_2)$ is the $i$-th homology group over $\mathbb{Z}_2$, that is $H_i(\mathbb{Z}_2)={\rm Ker}(\partial_i)/{\rm Im}(\partial_{i+1})$. In the next sections we discuss some concrete code families.


\subsection{2D Hyperbolic Surface Codes and Star Polyhedra}

Taking a surface ($D=2$), the only choice is for qubits to be associated with $1$-cells or edges so that $n=|E|$. We only consider regular tilings of the surface. Such tilings can be denoted by the Schl\"afli symbol $\{r,s\}$, meaning that each face is a regular $r$-gon and $s$ of such $r$-gons meet at each vertex. When $\{r,s\}$ is such that $\frac{1}{r}+\frac{1}{s}<\frac{1}{2}$, the surface is negatively curved or hyperbolic. For $\frac{1}{r}+\frac{1}{s}=\frac{1}{2}$, it is flat, and for $\frac{1}{r}+\frac{1}{s}>\frac{1}{2}$ it is positively curved, providing a regular tiling of the sphere. The last choice for $\{r,s\}$ gives us all the Platonic solids (e.g. the dodecahedron $\{5,3\}$) with trivial topology of the sphere, hence not interesting for encoding quantum information using topology since every closed loop can be contracted to a point. 

In order to make a code out of a hyperbolic surface, one needs to close the surface so it is topologically equivalent to a many-handled torus. The Euler characteristic $\chi$ of such tessellated closed surface equals $\chi=2-2g=|V|-|E|+|F|$ where $|V|$, $|E|$ and $|F|$ are the number of vertices, edges and faces and $g$ is the genus of the surface. The surface encodes $k=2g$ logical qubits. As was argued and reviewed in \cite{FM:surfacecode, BT:hyper}, hyperbolic surface codes based on an $\{r,s\}$-tiling have an encoding rate $\frac{k}{n}=1+\frac{2}{n}-2\left(\frac{1}{r}+\frac{1}{s}\right)$ and distance $d\geq c_{r,s} \log n$ with some constant $c_{r,s}$ which depends on the tessellation.
 
Some of the smallest codes that one obtains from this construction can be represented as uniform star polyhedra, see Table \ref{tab:star}. Examples are the dodedecadodecahedron based on closing a $\{5,4\}$-tiling of the hyperbolic plane \cite{Breuckmann+:hyper} with 60 qubits and the small stellated dodecahedron obtained from closing a $\{5,5\}$-tiling of the hyperbolic plane, depicted in Fig.~\ref{fig:Dodecahedron}. In its representation as star polyhedron, a regular 
$p$-gon can be represented as a star-$\frac{p}{k}$-gon ($k$ and $p$ mutually prime) whose vertices are generated by rotating by an angle $\frac{2\pi n p}{k}$ with integer $n$ \cite{book:coxeter}. The Schl\"afli-notation for a star polygon is $\{\frac{p}{k}\}$, that is, the pentagram is represented as $\{\frac{5}{2}\}$ so that the small stellated dodecahedron is denoted as $\{\frac{5}{2},5\}$. 


\begin{table}[htb]
\begin{tabular}{lc|c|c|c|c}
 & $n$ &  $k=2-\chi$ & $\text{wt}_Z$ ($\text{wt}_X$) & $d_Z$ ($d_X$) \\ 
\hline
Tetrahemihexahedron $U_{4}$ (a projective plane code) & 12 & 1 & 3,4 (4) &  3 (4) \\
Octahemioctahedron $U_{3}$ (a toric code) & 24 & 2 & 3,6 (4) & 4 (5) \\
Cubohemioctahedron $U_{15}$ (N.O.) & 24 & 4 & 4,6 (4) & 3 (4) \\
Small stellated dodecahedron $U_{34}$ (hyperbolic $\{5,5\}$) & 30 & 8 & 5 (5) & 3 (3) \\
Great dodecahedron $U_{35}$ (dual to $U_{34}$) & 30 & 8 & 5 (5) & 3 (3) \\
Small rhombihexahedron $U_{18}$ & 48 & 8 & 4,8 (4)  & 3 (4) \\ 
Small cubicuboctahedron $U_{13}$ & 48 & 6 & 3,4,8 (4) & 4 (4) \\
Great cubicuboctahedron $U_{14}$ & 48 & 6 &  3,4,8 (4) & 4 (4) \\
Great rhombihexahedron $U_{21}$ (N.O.) & 48 & 8 & 4,8 (4) & 3 (4) \\
Ditrigonal dodecadodecahedron $U_{41}$ (hyperbolic $\{5,6\}$) & 60 & 18 & 5 (6) & 3 (4) \cite{BT:hyper} \\
Small ditrigonal icosidodecahedron $U_{30}$ & 60 & 10 & 3,5 (6) & 4 (4)  \\
Great ditrigonal icosidodecahedron $U_{47}$ & 60 & 10 & 3,5 (6) & 4 (4) \\
Great dodecahemicosahedron $U_{65}$ & 60 & 10 & 5,6 (4) & 5 (4) \\
Small dodecahemicosahedron $U_{62}$ (N.O.) & 60 & 10 & 5,6 (4) & 5 (4) \\
Dodecadodecahedron $U_{36}$ (hyperbolic $\{5,4\}$) & 60 & 8 & 5 (4) & 6 (4) \cite{Breuckmann+:hyper}\\
Cubitruncated cuboctahedron $U_{16}$ & 72 & 6 & 6,8 (3) & 8 (4) \\
\end{tabular}
\caption{Some small uniform star polyhedra with $|E|=n$ physical qubits, $k=2g=2-\chi$ logical qubits, $Z$- (resp. $X$) parity check weight $\text{wt}_Z$ ($\text{wt}_X$) and $Z$- (resp. $X$) distance $d_Z$ and $d_X$. The distances $d_Z$ and $d_X$ were determined using the algorithm described in \cite{Breuckmann+:hyper}. A full list of uniform star polyhedra can be found at \url{https://en.wikipedia.org/wiki/List_of_uniform_polyhedra}. We omit all uniform polyhedra with $\chi=2$, all polyhedra with faces with 10 edges ($Z$-parity check weight 10) and all polyhedra with 120 physical qubits or more. N.O. indicates that the surface represented by the polyhedron is not orientable. Since each vertex looks the same (vertex-transitivity) in the polyhedron, all $X$-checks have the same, fairly low, weight and act the same. Except for the small stellated dodecahedron, all dual polyhedra have faces which are not regular polygons (they can be, say, arbitrary quadrilaterals), hence they are not star polyhedra. The many polyhedra with more than one type of polygonal face can also be viewed as quotient spaces of the uniformly-tiled hyperbolic plane.}
\label{tab:star}
\end{table}


\subsection{Some 3D and 4D Examples Based on Regular Tessellations}

If we consider regular tessellations of three-dimensional manifolds, we have the option of placing qubits on edges or faces. Since these are dual to each other, one can only construct one code from a given cell complex, so let's imagine that we associate qubits with faces.  Since 3D manifolds are complicated mathematical  objects, it is best to restrict any discussion to concrete three-dimensional cell complexes.  A honeycomb is a set of polyhedra filling space such that each face is shared by two polyhedra. We can use the Schl\"afli-symbol $\{p,q,r\}$ to denote a regular honeycomb, meaning that $r$ regular polyhedra, each of type $\{p,q\}$, meet at a vertex. There is only one regular honeycomb, namely $\{4,3,4\}$, a tiling by cubes, which fills flat 3D space and can be wrapped into a $3$-torus, hence leading to the 3D toric code.

The three-dimensional versions of the Platonic solids are 6 convex 4-polytopes: examples are $\{4,3,3\}$ (tesseract) and $\{5,3,3\}$ (120-cell) and its dual $\{3,3,5\}$ (600-cell). Instead of filling a flat space, these tile a sphere. In other words, similar as the dodecahedron is a regular tiling of the 2-sphere, one can view these cells as regular tilings (by volumes) of the 3-sphere $\mathbb{S}^3$. This implies that $\dim(H_2)=\dim(H_1)=0$, or no qubits are encoded in such objects. The Euler characteristic of these convex polytopes is $\chi(\mathbb{S}^3)=|V|-|E|+|F|-|C|=0$ (with $|C|$ the number of 3-cells). For example, the 120-cell $\{5,3,3\}$ has $|V|=600$, $|E|=1200$, $|F|=720$ and $|C|=120$.

Similar to the stellation of a dodecahedron, one can also stellate or greaten a 120-cell or a 600-cell to obtain so-called star polychora with non-trivial topology. An example is the small stellated 120-cell $\{\frac{5}{2},5,3\}$ which has $|F|=720$ qubits, $|C|=120$ $Z$-check cells, $|E|=1200$ $X$-check edges and $|V|=120$ vertices, hence its Euler characteristic is  $\chi=120-1200+720-120= -480$. Since $\chi=\sum_{i=0}^d (-1)^i \dim(H_i)$ and $\dim(H_0)=\dim(H_3)=1$, it follows that $-480=\dim(H_1)-\dim(H_2)$, so allowing for the encoding of logical information. \\


In 4D, a natural choice is to put qubits on 2-cells, so that one associates a $Z$-check with each 3-cell and an $X$-check with each edge. Beyond the 4D toric code which corresponds to a filling of flat 4D space, namely the honeycomb $\{4,3, 3, 4\}$ \cite{DKLP, BDMT:localdecoders}, generalizations of the hyperbolic surface codes to 4D are known to exist as well \cite{GL:hyperbolic4D, LL:golden}. These codes have a number of logical qubits~$k$ which scales linearly with the number of physical qubits~$n$, just like the hyperbolic surface codes. Unlike the 2D hyperbolic codes, the distance of these codes has been shown to scale polynomially with the number of physical qubits $d\in O(n^\epsilon)$ with $0<\epsilon<0.3$, see \cite{LL:golden}.
In principle one could create a code starting with a regular tessellation of 4D hyperbolic space by 4-polytopes. In order to have a closed 4D hyperbolic manifold one needs to find certain normal, torsion-free subgroups of the Coxeter group \cite{BT:hyper, thesis:breuckmann} such that 4-cells related by generators of this group can be identified. One known example is the orientable closed Davis manifold obtained from identifying opposing dodecahedra in the 120-cell, viewed as a 4-polytope \cite{R:foundations}.
It encodes $\dim(H_2)=72$ logical qubits and $n=144$ physical qubits (and $\dim(H_1)=\dim(H_3)=24$) \cite{RT:Davis}. 

In \cite{thesis:breuckmann} an exhaustive search for finding normal torsion-free subgroups of the $\{5,3,3,3\}$ tessellation of a 4D hyperbolic space is reported. In this tessellation qubits are associated with pentagons and dodecahedral cells act on 12 qubits. The $X$-checks correspond to tetrahedra in the dual lattice (with Schl\"afli-symbol $\{3,3\}$), having weight-4. Each qubit is acted on by 5 X-checks (${\rm deg}_X=5$) and 3 Z-checks (${\rm deg}_Z=3$). Unfortunately, running MAGMA to find an exhaustive list of small subgroups of this $\{5,3,3,3\}$  Coxeter group returns only one quantum code which encodes $k=197$ logical qubits ($\dim(H_2)=197$) into $n=16320$ physical qubits. For $\{5,3,3,3\}$ it is the only example which has less than $4\times 10^4$ physical qubits. 



\section{Features of The Small Stellated Dodecahedron Code}
\label{sec:features}

\begin{figure}[h]
\center
\begin{subfigure}[b]{0.3\textwidth}
            \centering
            \includegraphics[width=\textwidth]{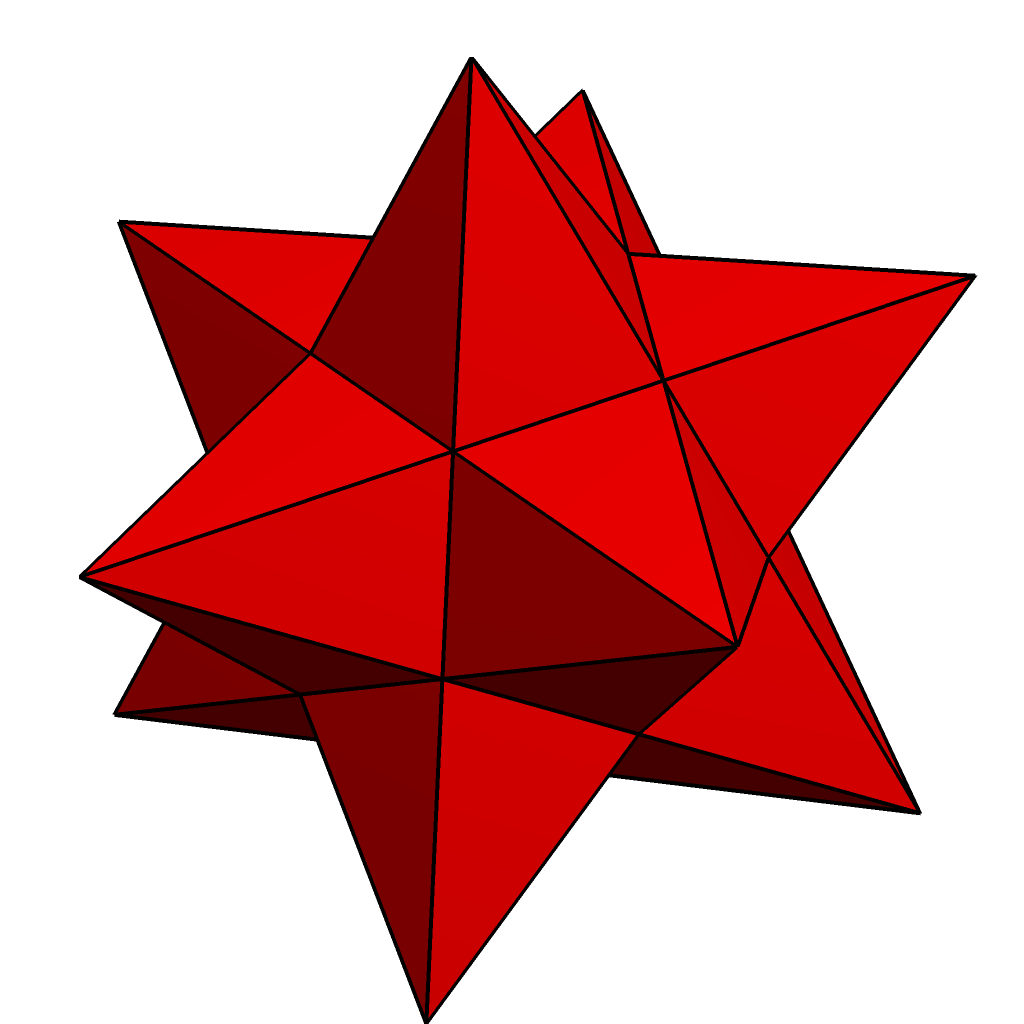}
            \caption{}
    \label{fig:SmallStellatedSub}
    \end{subfigure}
\begin{subfigure}[b]{0.28\textwidth}
            \centering
            \includegraphics[width=\textwidth]{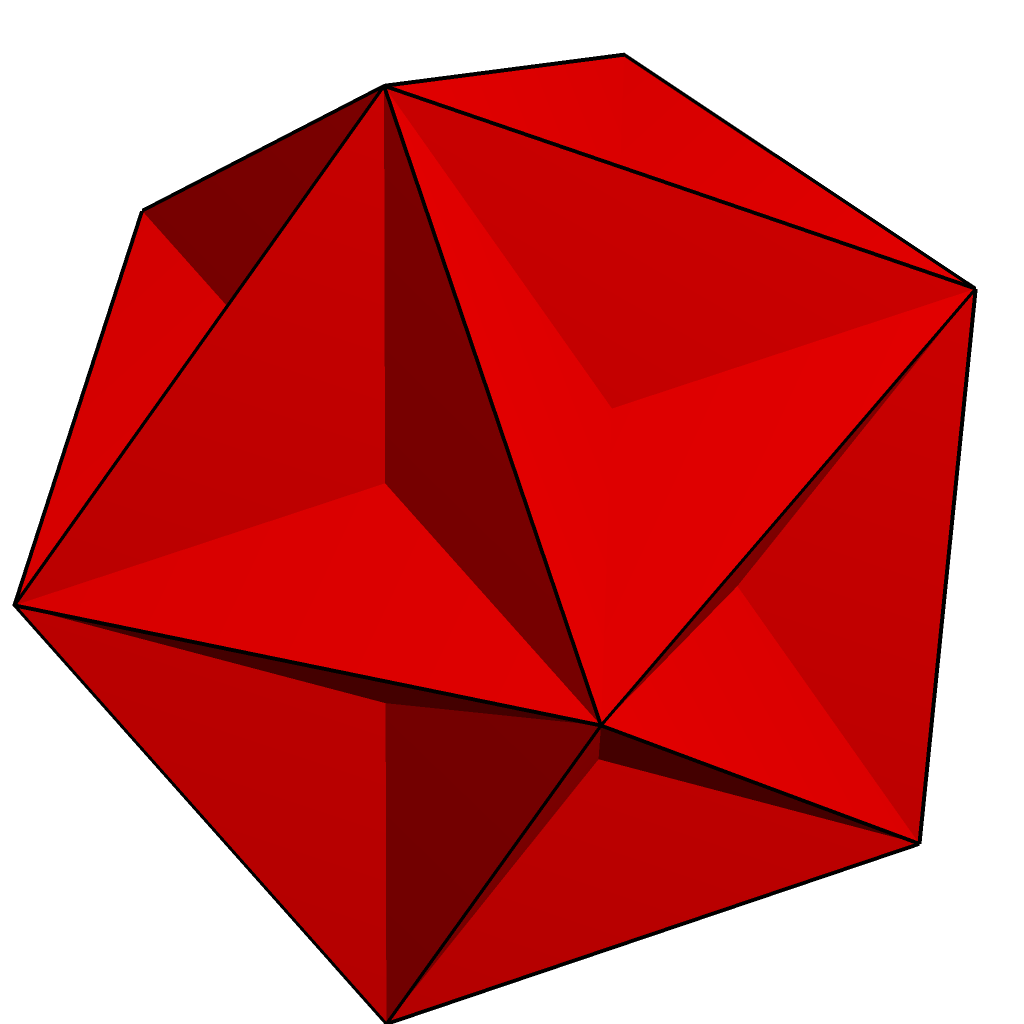}
            \caption{}
    \label{fig:GreatDodecSub}
    \end{subfigure}
\begin{subfigure}[b]{0.3\textwidth}
            \centering
            \includegraphics[width=\textwidth]{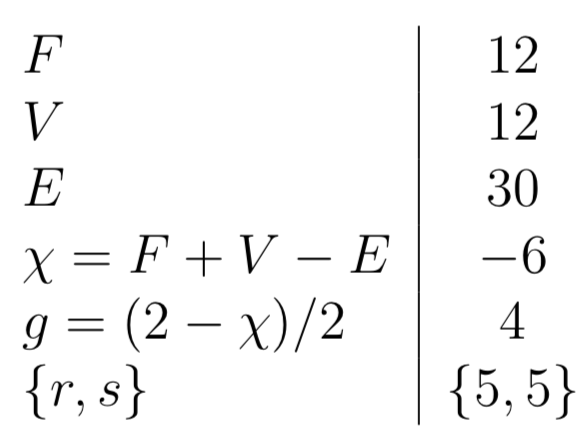}
            \caption{}
    \label{fig:TableGauss}
    \end{subfigure}
\caption{\small The small stellated dodecahedron as a $[[30,8,3]]$ code (\cref{fig:SmallStellatedSub}) and its dual polyhedron (\cref{fig:GreatDodecSub}) which is called the great dodecahedron. Both polyhedra have the same number of faces, vertices and edges. The vertices of the small stellated dodecahedron lie at the stars where edges meet. With the qubits placed on the edges, the Z-checks of the small stellated dodecahedron are described by intersecting pentagrammic faces, denoted as $\{\frac{5}{2}\}$. By computing the Euler characteristic $\chi$ using the parameters in \cref{fig:TableGauss}, it can be seen that the small stellated dodecahedron surface is topologically equivalent to the surface with genus $g=4$ \cite{rw:stella}.}
\label{fig:Dodecahedron}
\end{figure}


\begin{figure}[h]
	\centering
	\begin{minipage}{.4\textwidth}
		\centering
		\includegraphics[width=\linewidth]{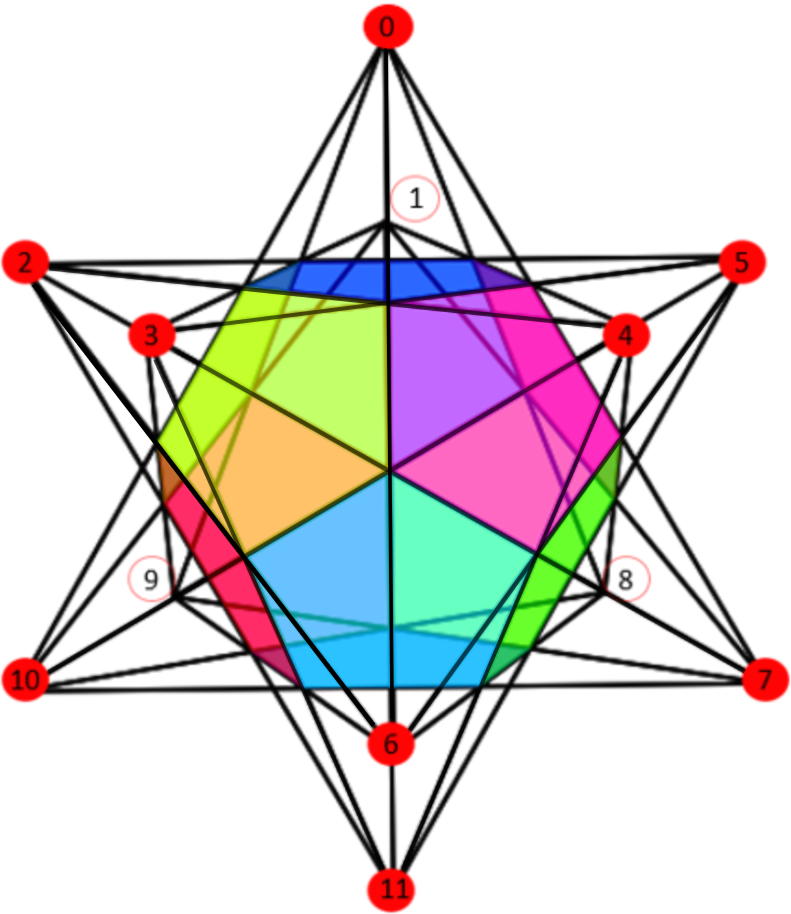}		
	\end{minipage}	
	\caption{Illustration of the small stellated dodecahedron construction by extending or stellating the edges of the (colored) regular dodecahedron until they intersect. The labeling of each vertex will be used to identify data qubits as well as the check and logical operators. For example, the check operator localized at vertex $0$ is $S^{X}_{0} = X_{(0,6)}X_{(0,7)}X_{(0,8)}X_{(0,9)}X_{(0,10)}$.}
	\label{fig:StellationLabels} 
\end{figure}


Some of the features of the small stellated dodecahedron code are summarized in \cref{fig:Dodecahedron}. The code encodes $8$ logical qubits (genus 4) into 30 and has distance 3. The $Z$-checks of the code are given by the pentagrammic faces, that is, a $Z$-check acts on the five edges of each pentagrammic face.
The $X$-checks are located at the vertices, i.e. an $X$-check acts on each of the five edges that meet at a vertex. 
There are thus 12 $X$- and $Z$-checks each of weight $\text{wt}(S)=5$. Since the product of all $X$-checks is $I$, the number of independent $X$-checks is 11 (and similarly there are 11 independent $Z$-checks). The small stellated docecahedron is obtained by stellating the dodecahedron as in \cref{fig:StellationLabels}, that is, we extend the edges until they meet at new vertices. One can understand the emergence of logical operators due to stellation for this specific polyhedron \footnote{For polyhedra one can also extend faces instead of edges, this is called greatening. An example is the greatening of the octahedron into the stella octangula. Since qubits are not defined on faces, this process does not create an interesting code. One can stellate the icosahedron into the small triambic icosahedron (which is dual to $U_{30}$, see Table \ref{tab:star}), but there does not seem to be an interpretation of stellation as creating non-trivial topology in this case.}.


The dodecahedron itself does not encode qubits but this trivial dodecahedron code has qubits on its 30 edges and weight-3 $X$ checks and weight-5 $Z$-checks which commute. Now we extend the edges, creating new vertices at which these edges meet. For this new code we keep the weight-5 dodecahedral $Z$-checks and add the weight-5 $X$-checks located at the 12 vertices where the extended edges meet. 
The 20 weight-3 $X$-checks of the dodecahedron still commute with the $Z$-checks and become possible logical operators.

\begin{figure}[h]
	\centering
	\begin{minipage}{.55\textwidth}
		\centering
		\includegraphics[width=\linewidth]{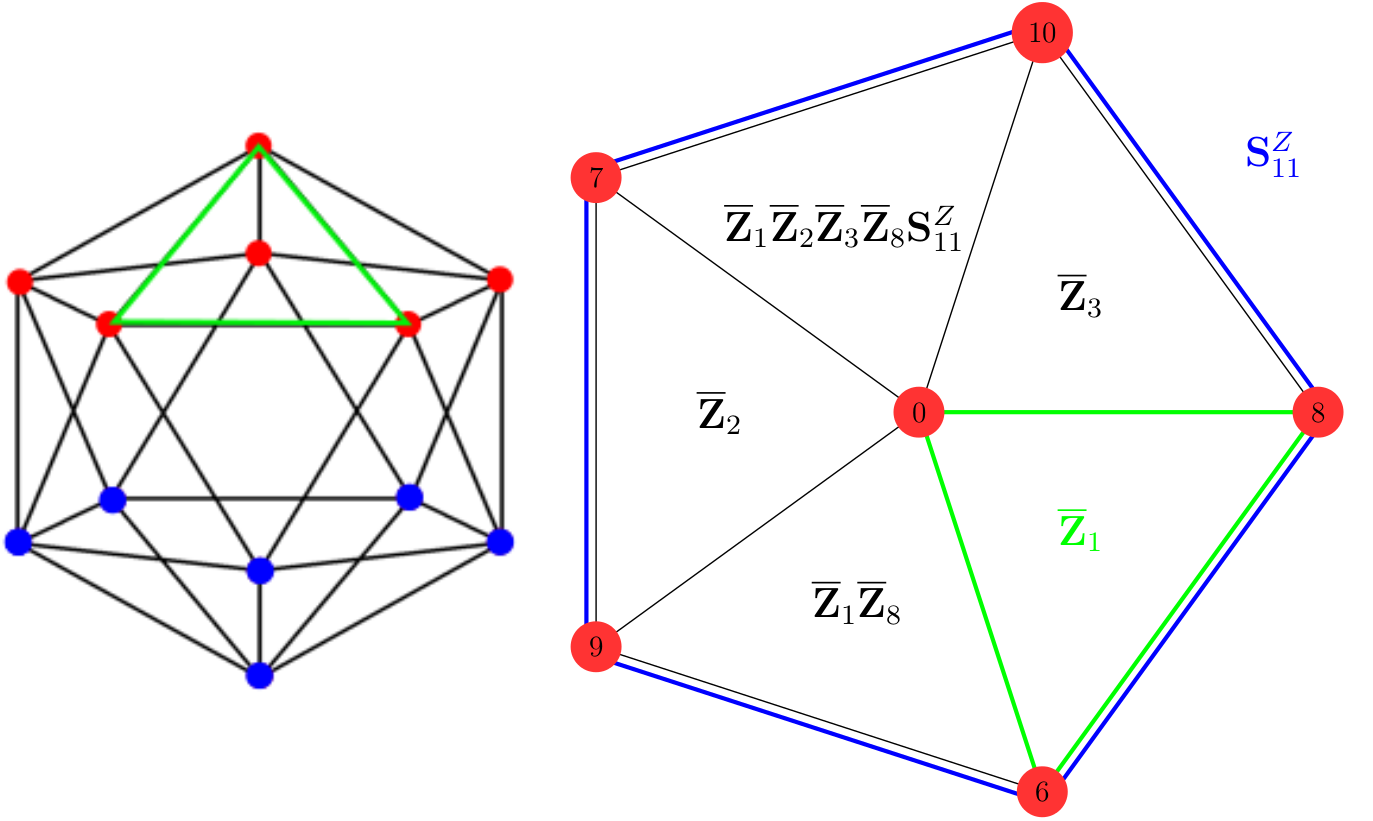}		
	\end{minipage}	
	\caption{Construction of the logical $Z$ operators ($\overline{Z}$) of the small stellated dodecahedron code from its disentangled graph, the icosahedron, where each blue or red vertex corresponds to an $X$-check acting on all incident edges. Each green triangle is a logical $Z$ operator commuting with all $X$-checks. In the right figure one sees how a product of 5 logical Z operators, listed in Table \ref{fig:StellationLabels}, equals the $Z$-check $S_{11}^Z$. $S_{11}^Z$ is the $Z$-check associated with the face located above the vertex labeled $11$ in \cref{fig:StellationLabels}.}
	\label{fig:DisentangledDodecLogZ} 
\end{figure}

In addition, the stellation process creates new weight-3 loops running along a triangle connecting three vertices and these loops cannot be the product of dodecahedral faces since the edges of the triangle lie in a single plane. When we take the 12 vertices and only represent the edges of the small stellated dodecahedron as a graph, one obtains an icosahedron, \cref{fig:DisentangledDodecLogZ}, which allows one to see the linear dependencies between the 20 $Z$-loops. In \cref{fig:DisentangledDodecLogZ}, the triangle logical $Z$-operators are represented by the highlighted green edges (any weight-two $Z$ operator would have odd support on at least one $X$-check). Around each vertex the product of 5 of these triangular $Z$-loops is a $Z$-check, hence the number of independent logical $Z$-operators is $20-12=8$. Similarly, the 20 vertices of the dodecahedron are logical $X$ operators, but products of 5 of them around a dodecahedral face are identical to one weight-5 $X$-check, so there are $20-12=8$ linearly-independent logical operators. A possible basis for the logical operators is given in \cref{tab:XZrepresentatives}.

\begin{table}[htb]
\begin{centering}
\begin{tabular}{|c|c|}
\hline 
Logical $\overline{Z}$s & Logical $\overline{X}$s  \tabularnewline
\hline 
$\overline{Z}_1 = Z_{(0,6)}Z_{(0,8)}Z_{(6,8)}$ & $\overline{X}_1 = X_{(0,6)}X_{(2,4)}X_{(3,5)}$\tabularnewline
\hline 
$\overline{Z}_2 = Z_{(0,7)}Z_{(0,9)}Z_{(7,9)}$ & $\overline{X}_2 = X_{(0,7)}X_{(1,4)}X_{(3,5)}$\tabularnewline
\hline 
$\overline{Z}_3 = Z_{(0,8)}Z_{(0,10)}Z_{(8,10)}$ & $\overline{X}_3 = X_{(0,10)}X_{(1,3)}X_{(2,4)}$\tabularnewline
\hline
$\overline{Z}_4 = Z_{(1,7)}Z_{(1,10)}Z_{(7,10)}$ & $\overline{X}_4 = X_{(1,7)}X_{(4,8)}X_{(5,11)}$\tabularnewline
\hline 
$\overline{Z}_5 = Z_{(2,5)}Z_{(2,6)}Z_{(5,6)}$ & $\overline{X}_5 = X_{(2,6)}X_{(3,11)}X_{(4,10)}$\tabularnewline
\hline 
$\overline{Z}_6 = Z_{(3,7)}Z_{(3,9)}Z_{(7,9)}$ & $\overline{X}_6 = X_{(2,4)}X_{(3,5)}X_{(3,7)}X_{(4,10)}$\tabularnewline
\hline 
$\overline{Z}_7 = Z_{(5,6)}Z_{(5,9)}Z_{(6,9)}$ & $\overline{X}_7 = X_{(1,11)}X_{(2,8)}X_{(5,9)}$\tabularnewline
\hline 
$\overline{Z}_8 = Z_{(0,8)}Z_{(0,9)}Z_{(6,8)}Z_{(6,9)}$ & $\overline{X}_8 = X_{(0,6)}X_{(0,10)}X_{(1,3)}X_{(1,11)}X_{(3,5)}X_{(5,11)}X_{(6,8)}X_{(8,10)}$\tabularnewline
\hline 
\end{tabular}
\par\end{centering}
\caption{\label{tab:XZrepresentatives}Set of independent logical $X$ and $Z$ operators of the small stellated dodecahedron code obeying 
$\overline{X}_i \overline{Z}_j=(-1)^{\delta_{ij}} \overline{Z}_j \overline{X}_i$. Each qubit is labeled by the edge $(u,v)$ with vertices $u,v$ ranging from 0 to 11 as in \cref{fig:StellationLabels}.}
\end{table}

The three-dimensional representation of this code as a small stellated dodecahedron immediately suggests (but does not necessitate) a placement and connectivity of qubits in 3D space. We have also argued in \cite{Breuckmann+:hyper} that any hyperbolic surface code can be implemented in a bilayer of qubits with CNOTs required between the two layers while the connectivity between qubits in each layer is planar. Recent experiments on superconducting qubits also demonstrate the feasibility of variable-range planar (hyperbolic) connectivity between qubits \cite{kollar+:hyper}.

\section{Parity Check Scheduling for LDPC CSS Codes}
\label{sec:ParityCheckSchedules}

In general, fault-tolerant error correction protocols for LDPC codes are implemented using entangling gates between ancilla and data qubits in order to measure the parity checks. In this section we assume that parity $X$-checks (resp. $Z$-checks) are measured via the interaction of a single ancilla qubit with $\text{wt}(X)$ data qubits via CNOT gates (resp. $\text{wt}(Z)$ data qubits via CNOT gates). Thus, at any point in time an ancilla qubit can interact via a CNOT with at most one data qubit. Similarly, any data qubit can interact with at most one ancilla qubit. A relevant problem is to find a scheduling of the CNOT gates which minimizes the number of time steps to measure all parity checks (so as to suppress the occurrence of errors).

This scheduling problem for homological codes based on non-flat geometries is not as trivial as it is for a surface code (or a 4D tesseract code \cite{DBT:4D}) where a local orientation and order in terms of North, East, South, West can be parallel transported over the whole lattice \cite{fowler+:review}. This idea does not translate to hyperbolic surface codes since the parallel transport of a vector around a closed curve does not bring it back to itself (in other words, the parallel transported vector depends on the path that one takes) capturing the local curvature. Hence we formulate the scheduling problem as an optimization problem which can be attacked numerically.

\begin{figure}[htb]
\center
\begin{subfigure}[b]{0.4\textwidth}
            \centering
            \includegraphics[width=\textwidth]{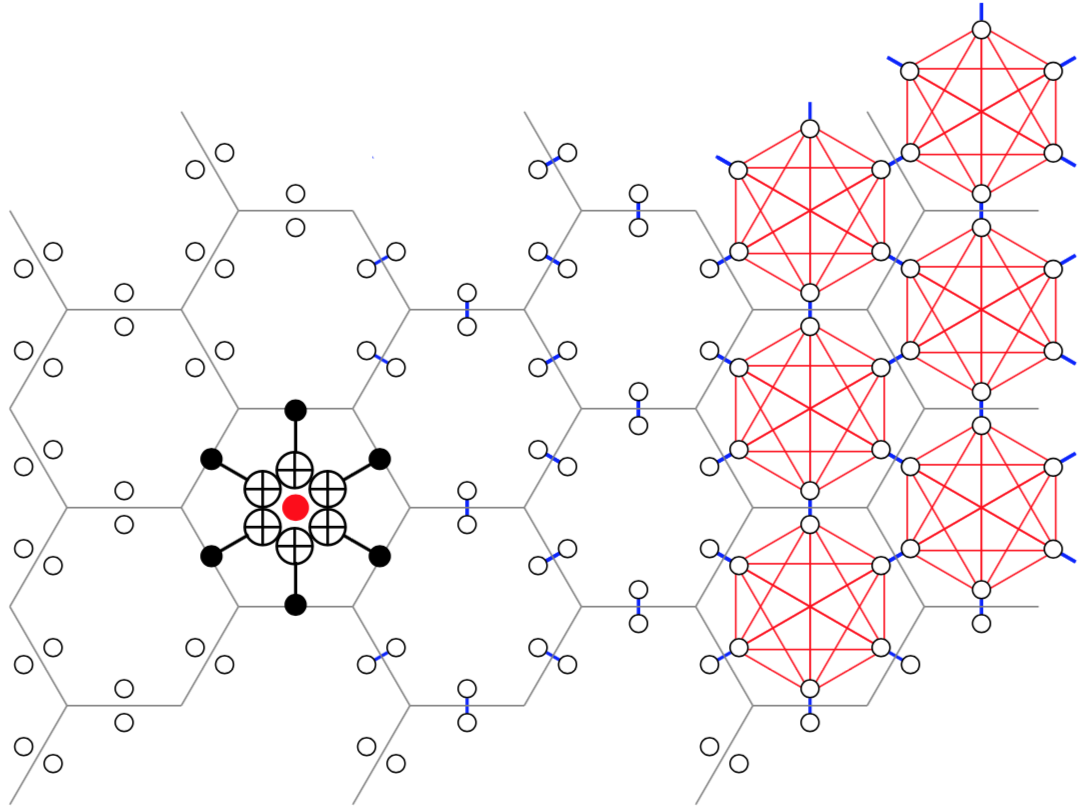}
            \caption{}
    \label{fig:FaceZGraphSched}
    \end{subfigure}
\begin{subfigure}[b]{0.32\textwidth}
            \centering
            \includegraphics[width=\textwidth]{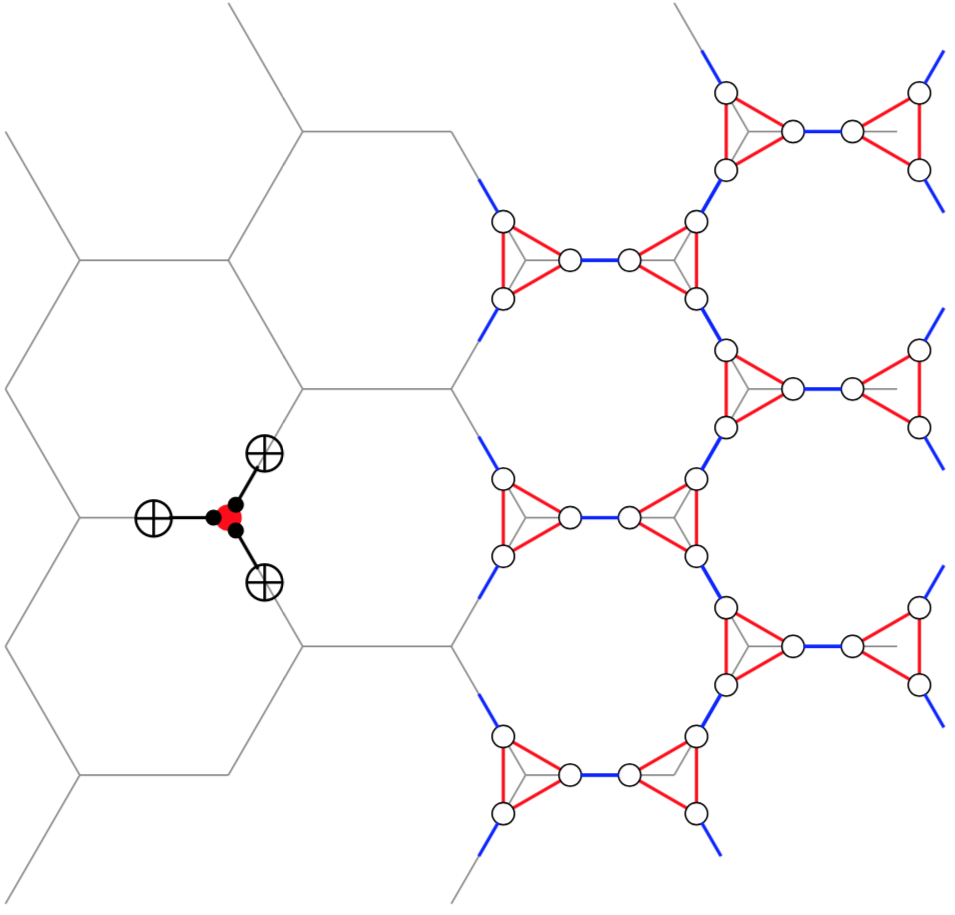}
            \caption{}
    \label{fig:VertexXGraphSched}
    \end{subfigure}
\caption{Separate $X$ and $Z$-scheduling graphs for the $\{ 6,3 \}$-tiling. In \cref{fig:FaceZGraphSched} the scheduling graph for $Z$-checks is shown whereas in \cref{fig:VertexXGraphSched} the scheduling graph for $X$-checks is shown. In \cref{fig:FaceZGraphSched} each qubit is replaced by two vertices connected by a blue edge. The vertices of all qubits which participate in a hexagonal face are connected by red edges.}
\label{fig:SchedGraph63Examp}
\end{figure}

For starters, let's imagine that we consider a CSS LDPC (low-density-parity-check) code and we wish to do all $X$-check measurements with maximal parallelism followed by an optimized schedule for the $Z$-check measurements. In such a scenario, the optimization of the number of time-steps in the measurement of all, say $X$-checks, corresponds to a graph vertex-coloring problem in a graph (to be defined) associated with the LDPC code. This graph and its coloring problem is (non-uniquely) obtained as follows. Each data qubit $q$ in the code is replaced by ${\rm deg}_X(q)$ vertices, together forming the vertex set $V_X$ of the $X$-check scheduling graph $G_X$. Here ${\rm deg}_X(q)$ is the number of $X$-checks that the qubit participates in, hence the number of CNOT gates that it has to undergo. The edges of $G_X$ are taken as follows. For each qubit $q$ we make a clique (complete subgraph) on its ${\rm deg}_X(q)$ vertices, capturing the constraint that none of the CNOT gates are simultaneous. For example, for homological surface codes, we replace each qubit by two connected vertices. Secondly, for each $X$-check of weight $\text{wt}(X)$ we create a complete graph $K_{\text{wt}(X)}$ between the vertices which represent the qubits on which the parity check acts, capturing the constraint that the CNOTs on the $X$-ancilla qubit cannot act simultaneously. Note that this choice is not unique as each qubit has ${\rm deg}_X(q)$ possible representatives. For homological surface codes, a natural choice which is the same for every edge and face is shown in \cref{fig:SchedGraph63Examp}. This gives the edge set $E_X$ of the scheduling graph $G_X=(V_X, E_X)$. 

Any vertex-coloring with $m$ colors of the graph $G_X$ gives a schedule which requires $T=m+2$ time-steps for the $X$-parity check measurements. In other words, the chromatic number $\chi(G_X)$ of the graph $G_X$ determines the number of required time steps. In the first time step, ancilla qubits are prepared in $\ket{+}$. In the subsequent $m$ time steps, CNOT gates are performed between data and ancilla qubits with the colors of vertices represented by the data qubits labeling the time step at which the CNOT is performed. Note that the coloring assignment only prescribes a temporal ordering up to permutations of time slots. In the last time step, the ancilla qubits are measured. A graph $G$ with maximum degree $\Delta(G)$ always admits a vertex coloring, i.e. $\chi(G) \leq \Delta+1$ \cite{book:bollabas}. The degree of $G_X$ is $\Delta(G_X)={\rm deg}_{X}+\text{wt}(X)-2$ when all qubits have the same degree ${\rm deg}_X$ and all parity checks have weight $\text{wt}(X)$. An example for a planar $\{6,3\}$-tiling is shown in \cref{fig:SchedGraph63Examp}. Note that for $\{r,s\}$-hyperbolic surfaces codes, the degrees of these scheduling graphs are $\Delta(G_X) =s$ and $\Delta(G_Z) = r$.

\begin{figure}[htb]
\center
\includegraphics[width=.5\textwidth]{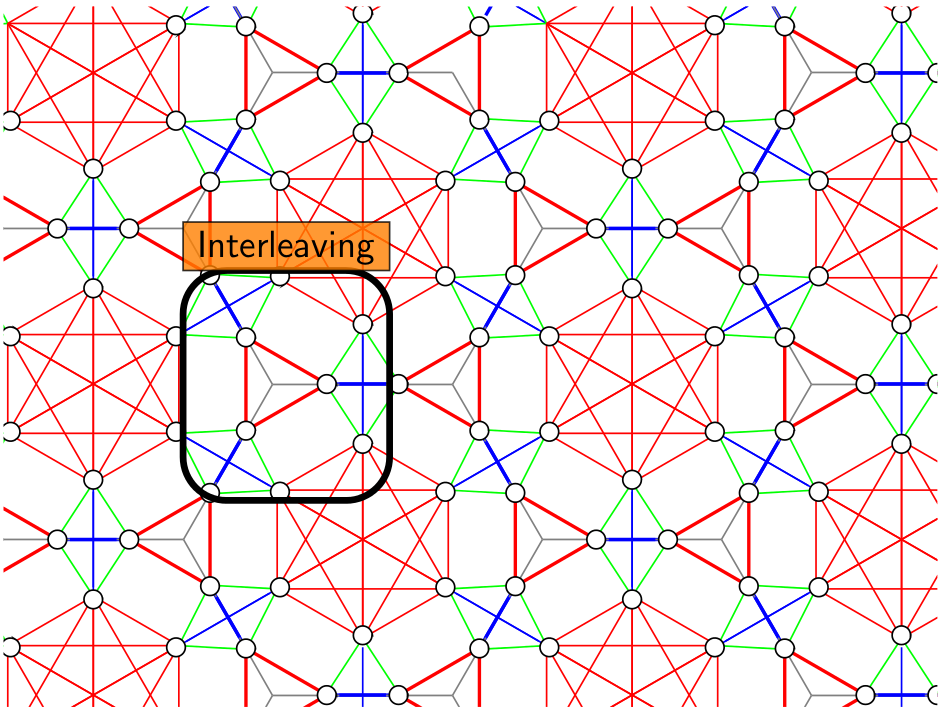}
\caption{The vertices and edges of the interleaved scheduling graph $G$ for a code based on a $\{6,3\}$-tiling: one takes the union of the vertices and edges in the graphs $G_X$ and $G_Z$ and adds additional green edges so that each qubit is represented by a clique of four vertices.
In the highlighted "interleaving" box, an $X$ and a $Z$-check act on the same pair of data qubits.}
\label{fig:ScheduleGraphAdditionalConstraint}
\end{figure}

However, in order to minimize the total number of time steps, it is advantageous to simultaneously apply CNOTs for $X$- and $Z$-check measurements instead of scheduling $X$ and $Z$-check measurements sequentially.  Such an {\em interleaved} schedule has been worked out for the surface code \cite{fowler+:review,TS:surface17} leading to a minimal schedule which requires $T=4+2$ time steps (including ancilla preparation and measurement).

Determining an optimal interleaved schedule can again be mapped onto a graph coloring problem obeying an additional constraint which ensures that there is no interference between the two types of measurements. To construct the interleaved scheduling graph $G$, we replace each qubit $q$ by ${\rm deg}_q$ vertices, where ${\rm deg}_q$ is now the total number of parity checks that the qubit participates in. This constitutes the set of vertices $V$. As to the edges, we again make each cluster of ${\rm deg}_q$ vertices into a clique. Then we add both the edges of the $X$- as well as the $Z$-checks as we did separately in the graph $G_X$ and $G_Z$.~\cref{fig:ScheduleGraphAdditionalConstraint} shows the example of the $\{6,3\}$-tiling. For codes with qubit degree ${\rm deg}$, $X$-parity checks of weight $\text{wt}(X)$ and $Z$-parity checks of weight $\text{wt}(Z)$, the degree of this interleaved scheduling graph equals $\Delta(G) ={\rm deg}-2+\max(\text{wt}(X), \text{wt}(Z))$. For $\{r,s\}$-surface codes, this results in $\Delta(G) =2+\max(r,s)$ so that $\chi(G) \leq 3+\max(r,s)$ since ${\rm deg}=4$. At the same time, the chromatic number $\chi(G) \geq \max(r,s)$ since the graph contains cliques of size $r$ and $s$. 

However, these coloring-based schedules may not be achievable since the CNOT order is additionally constrained due to the noncommutativity of Pauli $X$ and $Z$. In order to capture this constraint in the coloring problem, one can focus on homological surface codes in which $X$-checks and $Z$-checks have common support on either two or zero qubits (see also another expression of the constraints in \cite{LAR:colorcode}).

\begin{figure}[h]
\center
\begin{subfigure}[b]{0.32\textwidth}
            \centering
            \includegraphics[width=\textwidth]{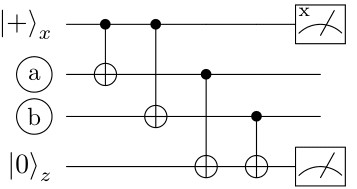}
            \caption{Proper circuit}
    \label{fig:prop}
    \end{subfigure}
\begin{subfigure}[b]{0.32\textwidth}
            \centering
            \includegraphics[width=\textwidth]{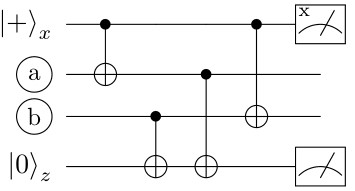}
            \caption{Improper circuit}
    \label{fig:improp}
    \end{subfigure}
\caption{The parity check circuits are proper when for each pair of qubits a  and b which are involved in a $X$- and a $Z$-check, both qubits first interact with the $X$-ancilla and then the $Z$-ancilla or both first with the $Z$-ancilla and then the $X$-ancilla.} 
\label{fig:CornerPropagation}
\end{figure}

Consider a pair of qubits, let's call them a and b, on which an $X$- and a $Z$-check both have support, see \cref{fig:CornerPropagation}. Both these qubits have to undergo CNOTs with an $X$-ancilla, as well as CNOTs with a $Z$-ancilla. Irrespective of what other data qubits are involved in the parity check measurements, the outcomes of the two measurements are {\em proper} when, either both qubits first interact with the $X$-ancilla and then with the $Z$-ancilla or vice versa. In these cases one can deduce (by propagating Pauli operators through the CNOT gates) that the measurement of $X_{x}$ of the $|+ \rangle_{x}$ ancilla equals the measurement of the observable $X_{x}X_{a}X_{b}I_{z}$. Since the $X$-ancilla is prepared in $\ket{+}_x$, this is equivalent to $X_{a}X_{b}I_{z}$. Similarly, a proper schedule shows that measurement of $Z_z$ is equivalent to measuring $I_{x}Z_{a}Z_{b}Z_{z}$, which is equivalent to $I_{x}Z_{a}Z_{b}$ due to the $Z$-ancilla being prepared in $\ket{0}_z$. For an improper circuit shown in \cref{fig:improp} the outcome of the parity checks is randomized since the $X$-check measurement depends on the expectation value of $X$ on $\ket{0}_z$ (and the $Z$-check depends on $Z$ on $\ket{+}_x$).


There is no general efficient algorithm to find the chromatic number of a graph since the problem is NP-complete.  However, for sparse graphs Ref. \cite{KY:coloring} (e.g. Theorem 5) discusses an efficient algorithm under some assumptions. However, our problem is compounded by the additional constraint that the schedule has to be proper. This means that a schedule of $5$ rounds of CNOT for the small stellated dodecahedron code might not be achievable, at least we have not found it. In addition, the schedule is required to be fault-tolerant which puts additional constraints on the schedule. For the small stellated dodecahedron code we have numerically obtained a sequential non-interleaved $X$ and $Z$-parity check schedule which is automatically proper.

We leave the existence of an efficient algorithm for determining a minimal-time parity check schedule for LDPC codes (with sufficiently large distance) as an open question.


\section{Fault-Tolerant Circuits for The Small Stellated Dodecahedron Code}
\label{sec:FaultTolerantSchemes}

In this section we present the fault-tolerant methods that will be used to analyze the performance of the small stellated dodecahedron code. The first step is to find a scheduling of the CNOTs used to measure the checks as discussed in the previous Section. By applying a degree of saturation (greedy) algorithm \cite{guide_col}, a separate schedule with 5 colors for both $G_X$ and $G_Z$ could be found (see \cref{fig:ScheduleAndColorResult,fig:FTcircuit}).
Consequently, all checks can be measured in $10+2$ time steps, \cref{fig:FTcircuit}. For this schedule we have verified that faults occurring during CNOT gates meet the requirements of fault-tolerance (see the discussion in \cref{sec:FT-details}).

We consider the following circuit-level depolarizing noise model for our analysis:
\begin{enumerate}
\item \label{it:1} With probability $p$, each two-qubit gate is followed by a two-qubit Pauli error drawn uniformly and independently from $\{ I,X,Y,Z \}^{\otimes 2}\setminus \{I \otimes I \}$.
\item \label{it:2} With probability $\frac{2p}{3}$, the preparation of the $| 0 \rangle$ state is replaced by $| 1 \rangle = X | 0 \rangle$. Similarly, with probability $\frac{2p}{3}$, the preparation of the $| + \rangle$ state is replaced by $| - \rangle = Z | + \rangle$.
\item \label{it:3} With probability $\frac{2p}{3}$, any single qubit measurement has its outcome flipped.
\item \label{it:4} Lastly, with probability $\frac{p}{10}$, each resting qubit location is followed by a Pauli error drawn uniformly and independently from $\{ X,Y,Z \}$.
\end{enumerate}

The reason to choose the idling location to have a lower error probability of $\frac{p}{10}$ is that it is a reasonable assumption for actual qubits (such as trapped-ion qubits \cite{bermudez+:ion} or nuclear spin qubits around a diamond NV center \cite{cramer+:QEC}) and it brings out more clearly the effect of CNOT errors which dominate the logical failure rate. Taking the idling location to have the same error probability $p$ as all other locations would give a disadvantage to the dodecahedron code versus the surface code since the parity check schedule for the dodecahedron code has more qubit idling.

As was shown in \cite{AGP06}  (see also the concise description in \cite{CB:flag}), a $d=3$ code should obey the following fault-tolerance criteria for an error correction (EC) unit in order that the logical error probability is possibly below the physical error probability $p$:

\begin{condition}{(Fault-Tolerant Criteria for an EC unit of a distance-3 code)}
\upshape
\begin{enumerate}
\item \label{it:1} If the input state has $r$ errors and the EC unit has $s$ faults with $r+s \le 1$, then ideal decoding of the output state of the EC will result in the same codeword as ideal decoding of the input state.
\item \label{it:2} Regardless of the number of errors in the input state, if there are $s$ faults during the EC unit with $s \le 1$, the output state can differ from a valid codeword by an error of at most weight $s$.
\end{enumerate}
\label{FTconditions}
\end{condition}
Here, ideal decoding means a round of fault-free error correction. Furthermore, by a fault we mean any gate, state-preparation, measurement or idle qubit failing according to the noise model described above. The second criteria states that if $E|\overline{\psi} \rangle$ is the input state with codeword $|\overline{\psi} \rangle$ and $\text{wt}(E)$ is arbitrary, the output state must be written as $E'|\overline{\phi} \rangle$ where $|\overline{\phi} \rangle$ is a codeword and $\text{wt}(E') \le s\leq 1$. Note that it is not required that $|\overline{\psi} \rangle = |\overline{\phi} \rangle$. 

\begin{figure}[htb]
\center
\includegraphics[scale=.4]{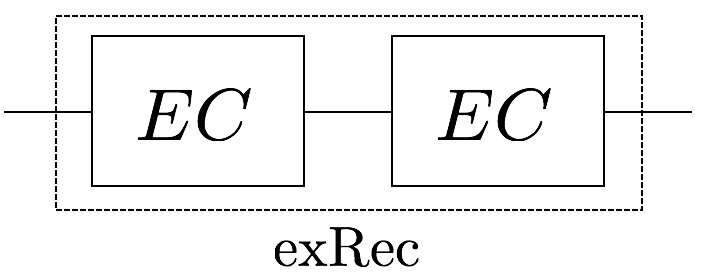}
\caption{Illustration of an extended rectangle (exRec). The EC unit consists of performing a round of fault-tolerant error correction (in our case, three rounds of syndrome measurements followed by the decoding protocol described in \cref{sec:FT-details}). The exRec consists of performing two consecutive EC's and its logical failure rate is determined by the occurrence of two malignant faults which lead to logical failure on output.}
\label{fig:ExRecFig}
\end{figure}

The second condition is particularly important in order to guarantee that errors won't accumulate during multiple rounds of EC resulting in a logical fault.  It was shown in \cite{AGP06} (and applied in e.g. \cite{CDT:study}) that it is the logical failure probability of an exRec, see \cref{fig:ExRecFig}, instead of the failure probability of a single EC unit that should be compared to the bare qubit failure probability $p$ in order to determine whether the lifetime of an encoded qubit will be longer than that of an unencoded qubit. The reason is that single faults in each consecutive EC unit can lead to logical failure since an incoming error and a fault in the unit can combine together. In the literature, pseudo-thresholds for small distance codes are often computed using a single EC unit. The pseudo-threshold is thus set by the total logical failure probability (probability of a logical $X$, $Y$ or $Z$ error) of the exRec being equal to $p$. In \cref{Sec:Numerics} we explicitly show that the logical failure rate of a single EC cannot be used to estimate the encoded qubit life-time.


\subsection{Decoder for $[[30,8,3]]$}
\label{sec:FT-details}

\begin{figure}[htb]
\center
\includegraphics[scale=.3]{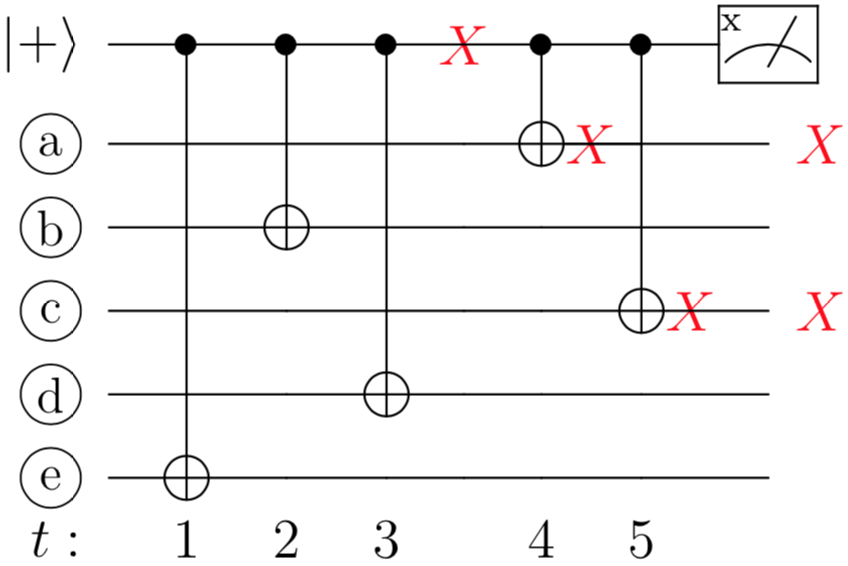}
\caption{Circuit for measuring a weight-five $X$-check. A single $X$ fault occurring after the third CNOT gate can propagate to two data qubits resulting in two outgoing $X$ errors.}
\label{fig:WeightFiveXstabCirc}
\end{figure}

Since the small stellated dodecahedron code is a distance-3 code, any single data qubit error in the EC unit will be corrected. However, the stabilizer checks are weight five which implies, as shown in \cref{fig:WeightFiveXstabCirc}, that a single fault occurring on some of the CNOT gates can lead to potentially dangerous errors of weight two. Note that for any check $P$ with $\text{wt}(P) = 5$, a single fault occurring during its measurement can lead to a data error $E$ with weight at most 2 since $\text{min}(\text{wt}(E),\text{wt}(EP)) \le 2$. Therefore, we need to ensure that any weight-two errors $E$ and $E'$ arising from a single fault during the measurement of an $X$ or $Z$ check either have a {\em unique syndrome} compared to each other ($s(E) \ne s(E')$) and compared to single faults which lead to an outgoing weight-one error, or if $s(E) = s(E')$, then they must be logically equivalent ($EE' \in \mathcal{S}$ where $\mathcal{S}$ is the stabilizer group). 

\begin{figure}[htb]
\center
\includegraphics[scale=.3]{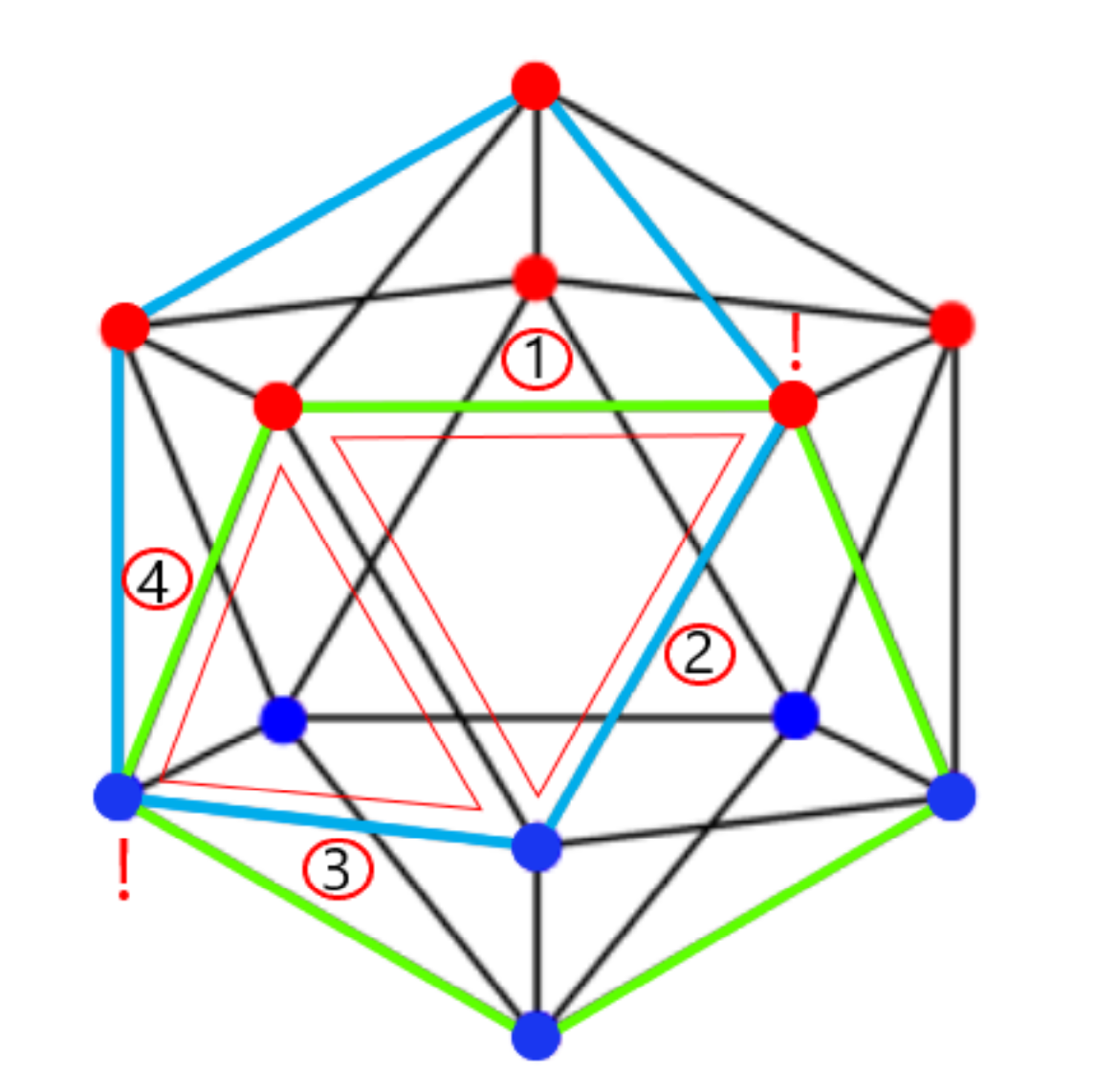}
\caption{Two pentagonal $Z$-checks, illustrated in green and blue, which overlap on two $Z$ logical operators (in red). The pairs of qubits, 1,4 and 2,3 are data qubits which can have $Z$ errors arising from a single fault occurring during the measurement of the two checks. The syndromes are indicated by the red exclamation marks.}
\label{fig:StabPlusLogicOverlap}
\end{figure}

A useful feature of the code is that the triangular logical $Z$ operators in \cref{fig:DisentangledDodecLogZ} overlap with any weight-5 $Z$-checks on at most 0 or 1 qubit: a triangular logical $Z$ lies in a plane which intersects the pentagrammic planes on at most one edge. However, an example of a problematic scenario involving a product of these logical operators is shown in \cref{fig:StabPlusLogicOverlap}. In this scenario, both pairs of qubits 1,4 and 2,3 could undergo $Z$ errors arising from a single fault during the measurement of the checks, and notice that both pairs of errors have the same error syndrome but {\em are not} logically equivalent. 
Thus, correcting the wrong error would lead to a logical fault. To resolve this issue, for the parity check schedule given in \cref{fig:ScheduleAndColorResult}, it was verified that every weight two $X$ or $Z$-error arising from a single fault during a stabilizer measurement has a unique syndrome. 


With the above considerations, we now describe a decoding protocol which satisfies the fault-tolerant criteria outlined in \cref{FTconditions}. Given the size of the small stellated dodecahedron code, it is possible to decode $X$ and $Z$ errors separately using full lookup tables (since each contains only $2^{11}=2048$ syndromes). For a given syndrome $s$, the lookup table chooses the lowest weight error $E$ that corresponds to the measured syndrome. However, note that there can be weight-two errors $E$ and $E'$ such that $s(E)=s(E')$ with $EE' \in N(\mathcal{S}) \setminus \mathcal{S}$ where $N(\mathcal{S})$ is the normalizer of the stabilizer group. Thus when constructing the lookup table, the corrections associated to all syndromes $s(E)$ where $E$ is a weight-2 error that can arise from a single fault during a stabilizer measurement should be $E$ and not some other weight-two error with the same syndrome as $E$ and which is not logically equivalent to $E$. Note that from the above discussion, this lookup table construction is possible. 

As with other distance-3 codes, a single round of syndrome measurement isn't sufficient to distinguish measurement errors from data qubit errors and would thus not be fault-tolerant. To make our decoding scheme fault-tolerant, we use the following protocol:

\vspace{17px}
\fbox{\begin{minipage}{35em}
        \textbf{\underline{Fault tolerant EC unit (for either $X$ or $Z$ errors)}:}
        
Perform three rounds of syndrome measurements resulting in syndromes $s_{1},s_{2}$ and $s_{3}$. Note that each round uses the CNOT scheduling depicted in \cref{fig:ScheduleAndColorResult} and \cref{fig:FTcircuit}.
\begin{enumerate}
   \item If at least two syndromes are trivial, apply no correction.
   \item If at least two syndromes $s$ are identical, apply the correction corresponding to $s$ using the lookup table.
   \item If the first two conditions are not satisfied, apply the correction corresponding to $s_3$ (the last measured syndrome) using the lookup table.
\end{enumerate}
\end{minipage}}

\vspace{17px}

Note that this procedure could be implemented to fault-tolerantly decode any distance-3 code, as long as one can pick a scheduling of the CNOTs (or other entangling gates) that guarantees that all errors arising from a single fault have unique syndromes (and those with the same syndrome are logically equivalent) and one uses these particular errors as the minimum weight corrections in the lookup table. 

We now give a rigorous proof that the above procedure satisfies the fault-tolerance criteria of \cref{FTconditions}. 

First, if there is an input error $E$ with $\text{wt}(E) = 1$ and no faults during the EC rounds, then all three rounds will report the syndrome $s(E)$ and the error will be corrected. Now suppose there are no input errors but a single fault occurs during the EC. If the fault occurs during the first round, then rounds two and three will produce the same syndrome and the resulting error will be corrected. If the fault occurs during the third round, then the first two rounds will yield a trivial syndrome and no correction will be applied. However, the output error must then be a correctable error. Thus ideally decoding the output would result in the input state. Now if the fault occurs during the second round, then all three syndromes could be different (depending at which time step the error occurred). There is also the possibility that $s_2=s_3$. In both cases, a correction corresponding to $s_3$ would be applied removing all errors on the data. Hence the first criterion will be satisfied.

Lastly, we need to show that the second criterion is satisfied. In fact, we modify the second criterion and demand that the output state differs from a valid codeword
by an error which is correctable by our ideal decoder (the ideal decoder is our Look-Up Table Decoder assuming no further errors). As discussed, this could be an error of weight-2. This modification does not alter the use of this condition in deriving fault-tolerance \cite{AGP06}.

In what follows, we will consider the case where the input error $E$ has arbitrary weight. If there are no faults during the EC, then all three syndromes will be equal to $s(E)$. Hence applying a correction $E'$ based on this syndrome will always project the code back to the code space (i.e. $EE' \in N(\mathcal{S})$). Now suppose there's a single fault during the first round of the EC. Then the syndromes $s_2=s_3$ will be the syndromes for the combined error $E$ and the resulting errors from the single fault during the first round. Thus correcting using $s_2$ will always project the code back to the code space. If the fault occurs during the second round, then, as in the previous paragraph, the correction will correspond to the last syndrome $s_3$ which includes both the input error and the error arising from the fault. Thus correcting using $s_3$ will always project the code back to the code space. Lastly, if the fault occurs during the third round, then the first two syndromes $s_1=s_2$ will correspond to the input error $E$. Let $E'$ be the resulting data qubit error from the third round. Then correcting using the recovery $\tilde{E}$ where $E\tilde{E} \in N(\mathcal{S})$, the output state will differ from a valid codeword by an error of weight $\text{wt(E')} \le 2$ which is correctable using our look-up table decoder.

\section{Numerical Results}
\label{Sec:Numerics}

In this section we present numerical results for the pseudo-thresholds of the surface-17 code of \cite{TS:surface17} and the small stellated dodecahedron code using the fault-tolerant decoding schemes and circuit-level noise model presented in \cref{sec:FaultTolerantSchemes}. To provide a fair comparison, we choose a sequential $X$ and $Z$-check schedule for the surface-17 code as well (such sequential schedule may be a necessity in some architectures anyhow, see e.g. the schedule in \cite{versluis:schedule}). Some of the code can be found at \url{https://github.com/einsteinchris}.

\begin{figure}[h]
\center
\begin{subfigure}[b]{0.5\textwidth}
            \centering
            \includegraphics[width=\textwidth]{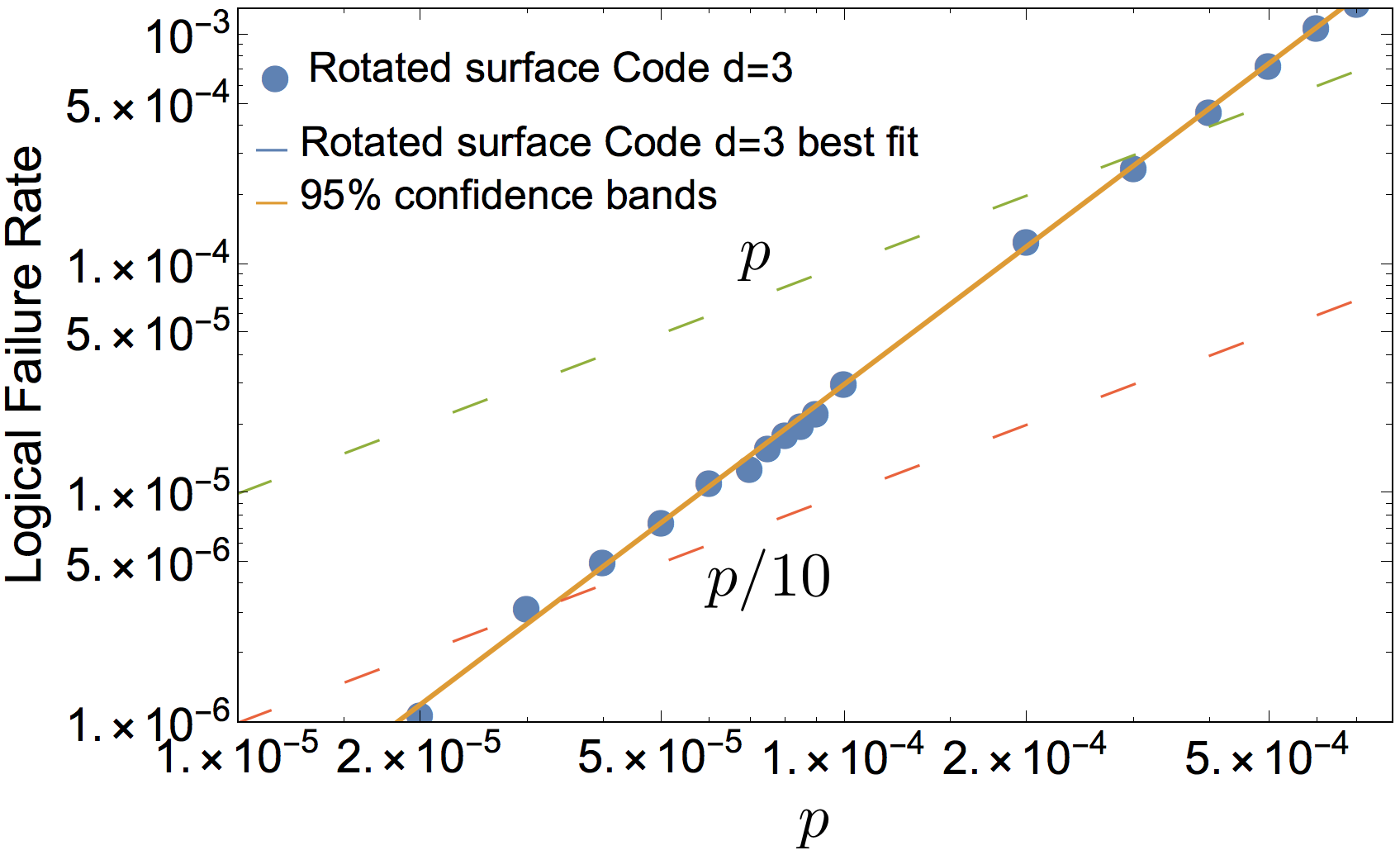}
            \caption{}
    \label{fig:Surface17Pseudo}
    \end{subfigure}
\begin{subfigure}[b]{0.48\textwidth}
            \centering
            \includegraphics[width=\textwidth]{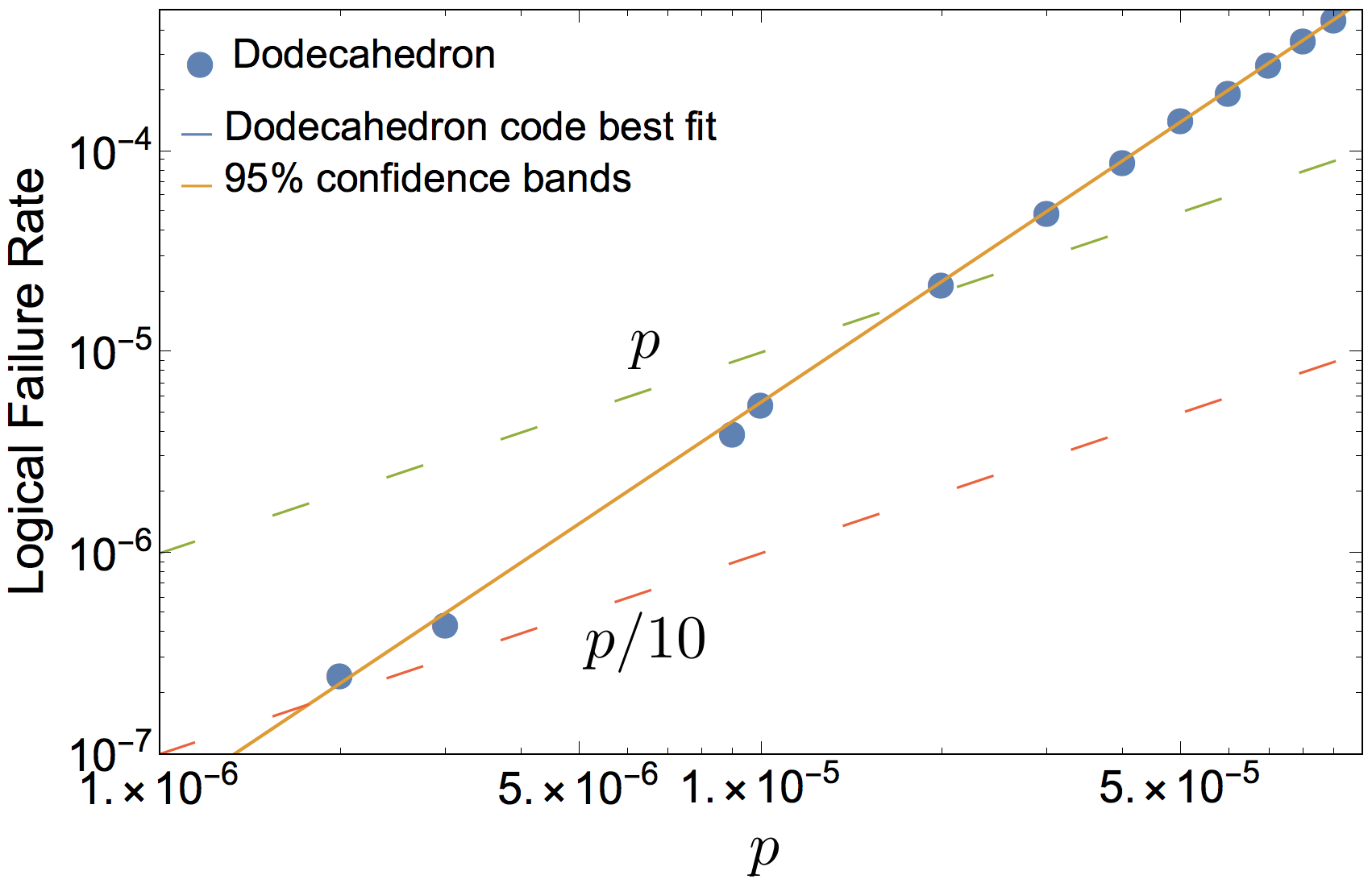}
            \caption{}
    \label{fig:DodecPseudo}
    \end{subfigure}
\caption{\cref{fig:Surface17Pseudo} and \cref{fig:DodecPseudo} show the total logical failure rate (probability of either an $X$, $Y$ or $Z$ logical fault) curve for the exRec of the Surface-17 ($p_L(p) \approx 3000p^{2}$) code and the exRec of the small stellated dodecahedron code ($p_L(p) \approx 56488p^{2}$). 	The intersection between these curves and the curve $f(p)=p$ gives, in principle, the pseudo-threshold of the codes. Note however that since idle qubits fail with probability $\frac{p}{10}$, for quantum memories, the relevant cross-over point is given by the intersection with the curve $\frac{p}{10}$ and not $p$. We find that it is $(3.32 \pm 0.01) \times 10^{-5}$ for Surface-17 and $(1.77 \pm 0.01) \times 10^{-6}$ for the small stellated dodecahedron code.}
\label{fig:pseudoThresholdsSurfaceDodec}
\end{figure} 



\begin{figure}[h]
\center
\begin{subfigure}[b]{0.49\textwidth}
            \centering
            \includegraphics[width=\textwidth]{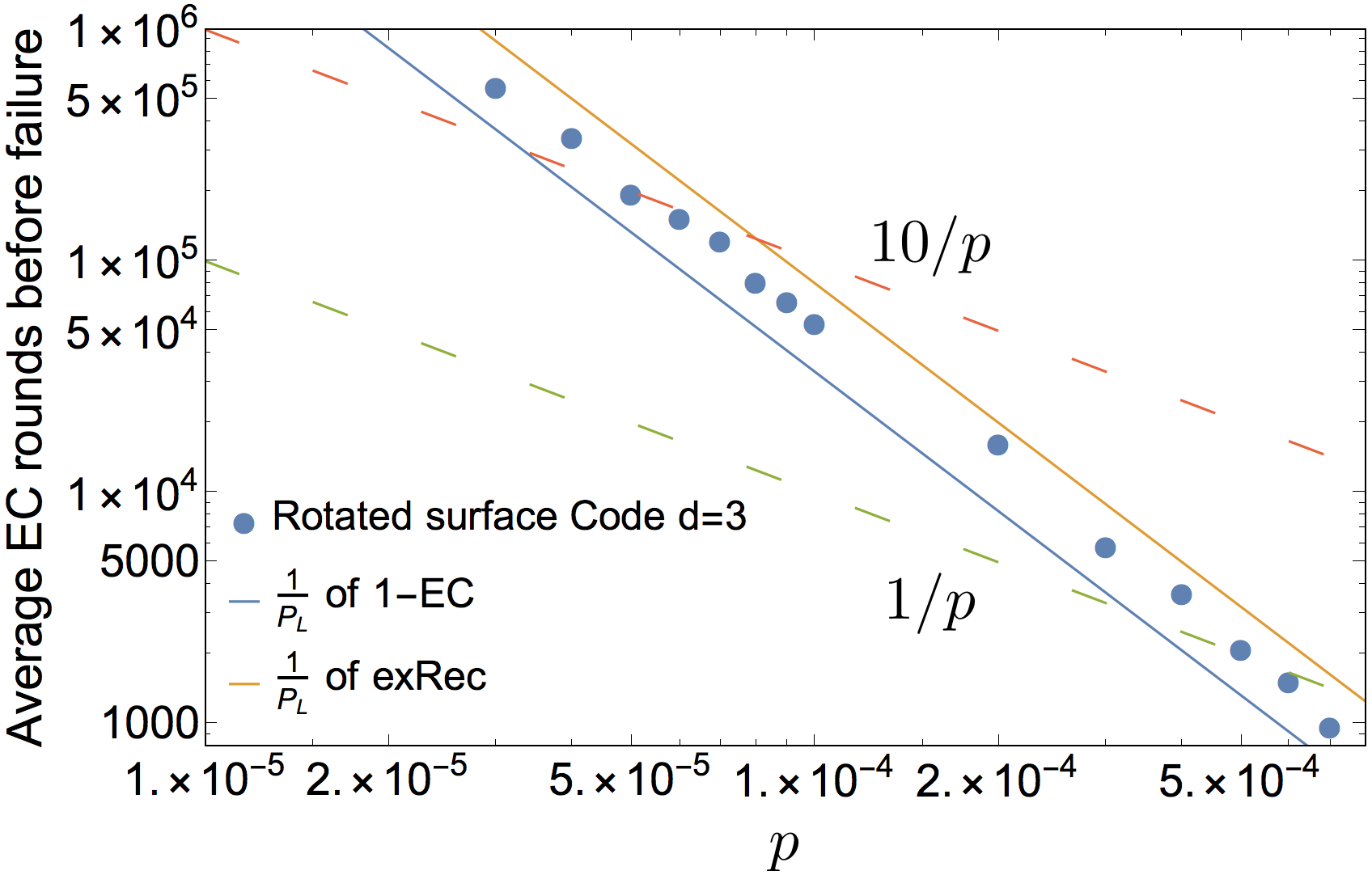}
            \caption{}
    \label{fig:SurfaceCodeLifetime}
    \end{subfigure}
\begin{subfigure}[b]{0.49\textwidth}
            \centering
            \includegraphics[width=\textwidth]{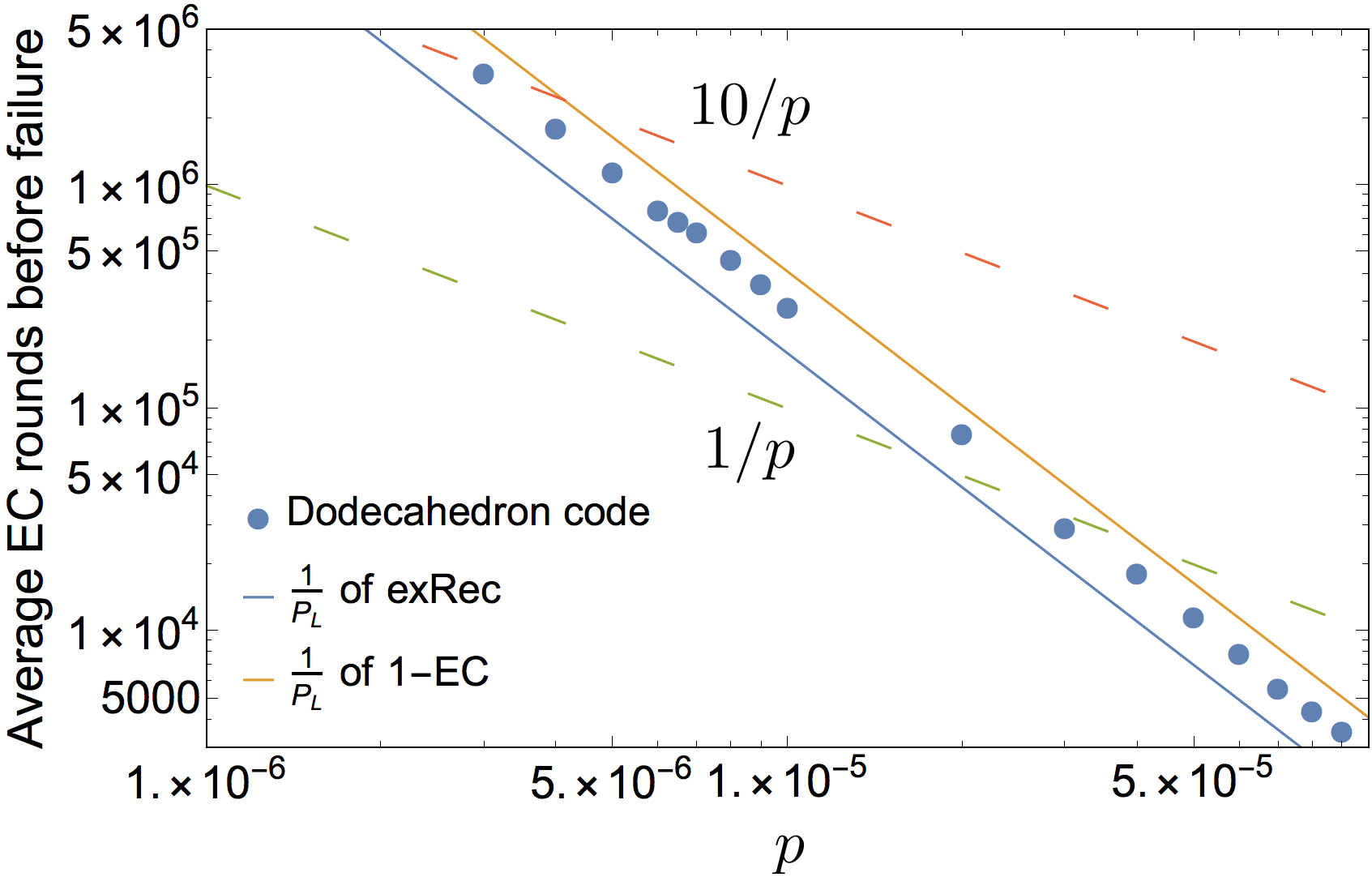}
            \caption{}
    \label{fig:DodecLifeTime}
    \end{subfigure}
\caption{\cref{fig:SurfaceCodeLifetime} shows the average number of EC rounds before failure of an encoded qubit in the surface-17 code while \cref{fig:DodecLifeTime} shows the average number of EC rounds before failure of 8 encoded qubits in the small stellated dodecahedron code. Solid lines show $1/p_L$ where $p_L$ is the logical failure rate (as a function of $p$) obtained for both the exRec circuit and the single EC unit circuit. The data clearly shows that the lifetime is lower bounded by $1/p_L$ obtained from the exRec circuit and not a single EC unit.}
\label{fig:LifeTimePlots}
\end{figure}

To obtain the average lifetime of a physical qubit, suppose that an error is introduced with probability $p$ at any given time step. The probability that an error is introduced after exactly $t$ time steps is given by $f_p(t) = (1-p)^{t-1}p$. Thus the mean time before a failure occurs is given by $\sum_{t=1}^{\infty} tf_p(t) = 1/p$. To obtain a lower bound of the lifetime of an encoded qubit, we can simply replace $p$ by the logical failure rate curve $p_L(p)$ of the exRec (see \cite{AGP06}). 
For a distance-3 code, $p_L(p) = cp^{2} + O(p^3)$ since the code can correct any single qubit error.

In \cref{fig:pseudoThresholdsSurfaceDodec}, plots illustrating the pseudo-threshold of the Surface-17 and the small stellated dodecahedron code are shown. In \cref{fig:LifeTimePlots}, the circular dots show the average number of EC rounds before failure of encoded qubits for both a single qubit encoded in Surface-17 and 8 qubits encoded in the small stellated dodecahedron code (in the simulation, we decoded every three rounds and propagated residual errors into the next EC unit).  Unfortunately, the Surface-17 code has a pseudo-threshold which is about 19 times larger than the dodecahedron code ($(3.32 \pm 0.01) \times 10^{-5}$ compared to  $(1.77 \pm 0.01) \times 10^{-6}$).
 Note that the pseudo-threshold values were obtained by the intersection between the curve $\frac{p}{10}$ (since we are considering a noise model where idle qubits fail with probability $\frac{p}{10}$ and are concerned about quantum memories) and the logical failure rate curve of the exRec. 
 The differences in pseudo-thresholds are primarily due to the larger number of locations in the fault-tolerant circuits of the dodecahedron code compared to the surface-17 code circuits as well as the fact that both codes have the same distance. In fact, just by counting the number of pairs of CNOT gates in an EC unit one can get an indication of the pseudothreshold. For the small stellated dodecahedron code the number of CNOT gates is $3 \times 5 \times 22=330$ so that ${330 \choose 2}=54285$ while for Surface-17 the number of CNOT gates in an EC-unit is $3 \times 4 \times 8=96$ so that ${96 \choose 2}=4560$, in rough correspondence with the $c$'s in $P_L(p)=c p^2$ observed in \cref{fig:pseudoThresholdsSurfaceDodec}.

Note that since the dodecahedron code encodes 8 logical qubits, a fairer comparison would be to compare the logical failure rate of the dodecahedron code with that of 8 qubits encoded in the surface-17 code. In general, if the logical failure rate of an extended rectangle of the code is given by $p_L(p)$, the logical failure rate of $m$ copies of the code is given by  $p_L^{(m)}(p) = 1-(1-p_L(p))^{m} = m p_L(p) + \mathcal{O}((p_L(p))^2)$.

\begin{figure}[h]
\center
\includegraphics[scale=.3]{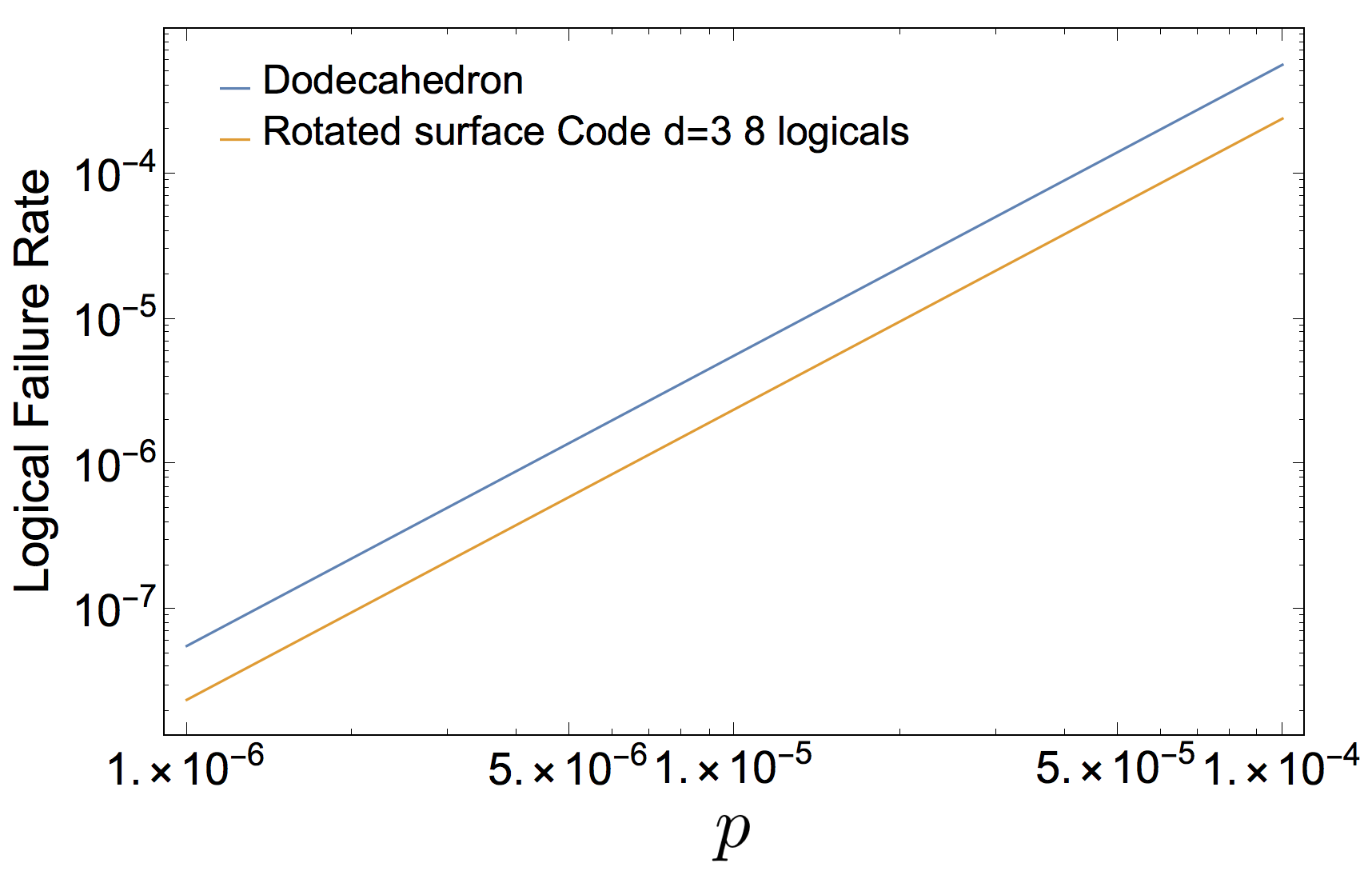}
\caption{Comparison of eight logical qubits encoded in the surface-17 code, with a total logical failure rate given by $1-(1-p_L(p))^{8}$ where $p_L(p) \approx 3000p^{2}$, with 8 logical qubits encoded in the dodecahedron code with $p_L(p) \approx 56488p^{2}$. It can be seen that the surface-17 still outperforms the dodecahedron code since $8 \times 3000 \ll 56488$.}
\label{fig:Compare8logicalSurface1logicalDodec}
\end{figure}

In \cref{fig:Compare8logicalSurface1logicalDodec}, we compare the logical failure rate of eight qubits encoded in the dodecahedron code with eight qubits encoded in the surface-17 code. It can be seen that the surface-17 code still achieves a lower logical failure rate compared to the dodecahedron code. 

\section{Discussion}
The fault-tolerance analysis for the small stellated dodecahedron code has shown the difficulty of getting a block code with high pseudo-threshold when we include noise in the parity check circuits themselves. The EC unit of this code is simply larger since many more checks need to be measured and the pseudo-threshold is determined by pairs of malignant locations in this large unit. In contrast, separate copies of the surface code, each with a much smaller EC unit benefit from having "room for each logical operator". One might expect that this problem becomes less severe for larger hyperbolic codes which have shown lower but still good performance versus surface codes for a phenomenological noise model \cite{Breuckmann+:hyper}. 

One could consider how Steane error correction can improve the performance of the small stellated dodecahedron code: we expect that the qubit overhead will be larger (mainly due to the requirement for preparing four logical $\ket{0}$ and four logical $\ket{+}$ ancillas) but the pseudo-threshold would be quite better. 
The tetrahemihexahedron code $[[12,1,3]]$ (see Table \ref{tab:star}) with some weight-3 checks might be an interesting variation on the $3 \times 4$ rotated surface code (with $d_Z=3, d_X=4$).

Lastly, we also tried to use only four of the eight logical qubits of the small stellated dodecahedron code for encoding logical information in order to see if significant improvements in the pseudo-threshold could be obtained. However, our numerical simulations showed that for various choices of the logical qubits, the pseudo-threshold improved by less than a factor of two. The primary reason is that in most cases where a failure occurred, several logical qubits were afflicted. 

A goal for future work would be to compare the performance of the small stellated dodecahedron code with the surface code for a physically-motivated noise model in an optically-linked ion-trap architecture \cite{nigmatullin+:ion} or an optically-linked NV-center in diamond architecture \cite{vanDam+:entanglement}. 
\enlargethispage{20pt}





\funding{We acknowledge support through the EU via the ERC GRANT EQEC No. 682726. This research was supported in part by Perimeter Institute for Theoretical Physics. Research at Perimeter Institute is supported by the Government of Canada through Industry Canada and by the Province of Ontario through the Ministry of Economic Development $\&$ Innovation. C.C. acknowledges the support of NSERC through the PGS D scholarship.}

\ack{We would like to thank Kasper Duivenvoorden and Christophe Vuillot for useful discussions. We acknowledge the use of valuable computing time on the RWTH Aachen Compute Cluster. We thank Koen Bertels for quick access to the 4 machines at Computer Engineering TU Delft, and Steve Weiss for the use of computing clusters at IQC Waterloo. C.C. would like to acknowledge TU Delft for its hospitality where the work was completed.}

\bibliographystyle{RS}

\bibliography{homological}

\appendix

\section{Coloring and Parity Check Circuits for The Small Stellated Dodecahedron Code}
\label{app}

\begin{figure}[htb]
	\center
   	\centering
    \includegraphics[width=.55\textwidth]{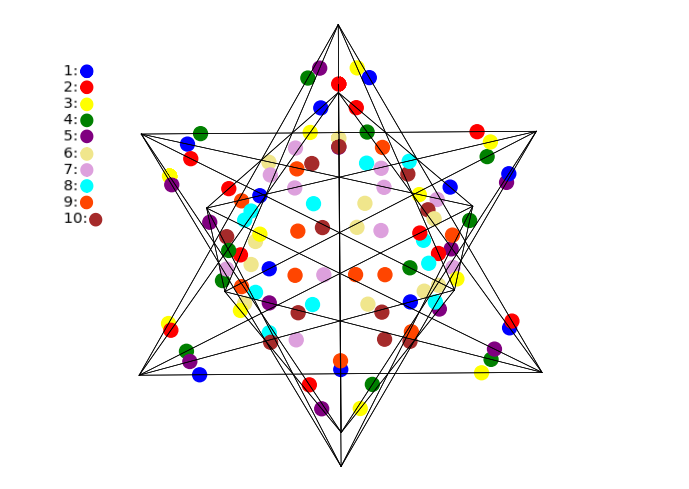}        
    \label{fig:ColoringResult}
    \caption{A coloring with 10 colors, see a corresponding schedule in Fig.~\ref{fig:FTcircuit}.}
\label{fig:ScheduleAndColorResult}
\end{figure}

\begin{figure}[htb]
\center
  \includegraphics[width=\textwidth]{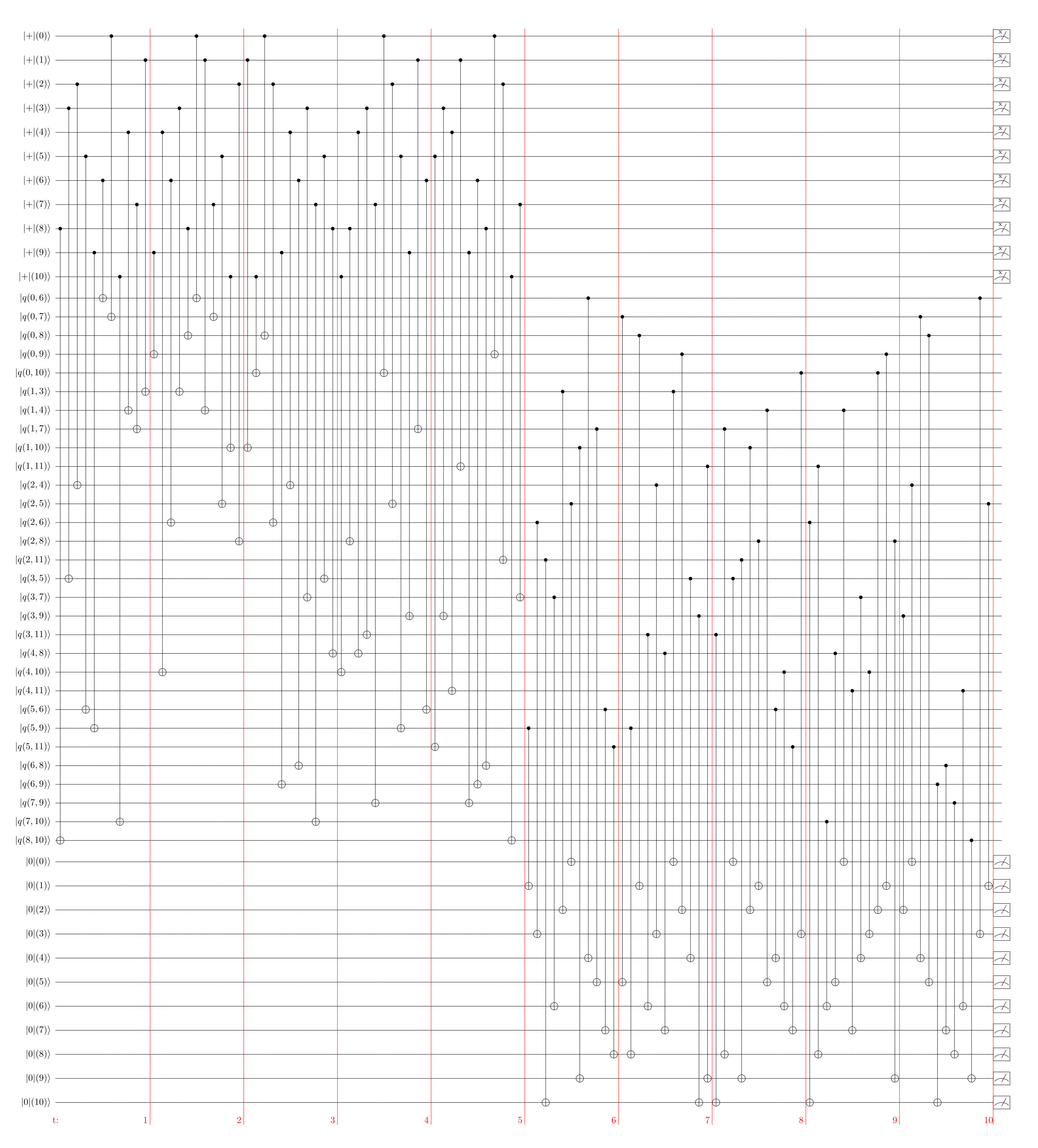}
  \caption{The parity check circuit for one EC round for the small stellated dodecahedron code involving $11$ ancilla qubits starting in $\ket{+}$ for the $X$-check measurement and $11$ ancilla qubits in $\ket{0}$ for the $Z$-check measurement. The parity check measurements are accomplished in 10 rounds (separated by red vertical lines), corresponding to a graph coloring with 10 colors. }
\label{fig:FTcircuit}
\end{figure}


\end{document}